\documentstyle[11pt]{article}
\renewcommand{\baselinestretch}{1.3}

\input epsf                            

\begin{document}
%\sloppy
%\input{pref}
\newcount\nummer \nummer=0
\def\f#1{\global\advance\nummer by 1 \eqno{(\number\nummer)}
      \global\edef#1{(\number\nummer)}}
%begin macros

% ohne Labels
\newcommand{\pref}[1]{\ref{#1}}
\newcommand{\plabel}[1]{\label{#1}}
\newcommand{\prefeq}[1]{Gl.~(\ref{#1})}
\newcommand{\prefb}[1]{(\ref{#1})}
\newcommand{\prefapp}[1]{Appendix~\ref{#1}}
\newcommand{\plititem}[1]{\begin{zitat}{#1}\end{zitat}}
\newcommand{\pcite}[1]{\cite{#1}}
\newcommand{\plookup}[1]{\hoch{\ref{#1}}}
% ohne Labels

\def\Figtwelve{12}
\def\Figthirteen{13}
\def\trace{\mbox{tr}\,}
\let\oe=\o
\def\attn{{\Large $\bigtriangledown
  \mbox{\scriptsize\hspace{-.276cm}!}$\hspace{.1cm}}}
\def\Di{\displaystyle}
\def\nn{\nonumber \\}
\def\be{\begin{equation}}
\def\ee{\end{equation}}
\def\ba{\begin{eqnarray}}
\def\ea{\end{eqnarray}}
\def\la{\plabel} \def\pl{\label}
\def\re{(\ref }%\def\pr{(\pref}
\def\rz#1 {(\ref{#1}) }   \def\ry#1 {(\ref{#1})}
\def\el#1 {\plabel{#1}\end{equation}}
\def\rp#1 {(\ref{#1}) }
\def\i{{\rm i}}
\let\a=\alpha \let\b=\beta \let\g=\gamma \let\d=\delta
\let\e=\varepsilon \let\ep=\epsilon \let\z=\zeta \let\h=\eta \let\th=\theta
\let\dh=\vartheta \let\k=\kappa \let\l=\lambda \let\m=\mu
\let\n=\nu \let\x=\xi \let\p=\pi \let\r=\rho \let\s=\sigma
\let\t=\tau \let\o=\omega \let\c=\chi \let\ps=\psi
\let\ph=\varphi \let\Ph=\phi \let\PH=\Phi \let\Ps=\Psi
\let\O=\Omega \let\S=\Sigma \let\P=\Pi \let\Th=\Theta
\let\L=\Lambda \let\G=\Gamma \let\D=\Delta

\def\wt{\widetilde}
\def\mytilde{\widetilde} % ausser fuer CQG
\def\w{\wedge}
\def\0{\over } \def\1{\vec } \def\2{{1\over2}} \def\4{{1\over4}}
\def\5{\bar } \def\6{\partial }
\def\7#1{{#1}\llap{/}}
\def\8#1{{\textstyle{#1}}} \def\9#1{{\bf {#1}}}

\def\({\left(} \def\){\right)} \def\<{\langle } \def\>{\rangle }
\def\lb{\left\{} \def\rb{\right\}}
\let\lra=\leftrightarrow \let\LRA=\Leftrightarrow
\let\Ra=\Rightarrow \let\ra=\rightarrow
\def\ul{\underline}
                          
\let\ap=\approx \let\eq=\equiv % \let\ex=\times \let\hc=\dagger
\let\ti=\tilde \let\bl=\biggl \let\br=\biggr
\let\bi=\choose \let\at=\atop \let\mat=\pmatrix
\def\CL{{\cal L}}\def\CX{{\cal X}}\def\CA{{\cal A}}
\def\CF{{\cal F}} \def\CD{{\cal D}} \def\rd{{\rm d}}
\def\rD{{\rm D}} \def\CH{{\cal H}} \def\CT{{\cal T}} \def\CM{{\cal M}}
\def\CI{{\cal I}} \newcommand{\dR}{\mbox{{\rm I \hspace{-0.86em} R}}}
  \newcommand{\dN}{\mbox{{\rm I \hspace{-0.865em} N}}}
\def\CP{{\cal P}}\def\CS{{\cal S}}\def\C{{\cal C}}
%end macros

\begin{titlepage}
\renewcommand{\thefootnote}{\fnsymbol{footnote}}
\renewcommand{\baselinestretch}{1.3}
\hfill  TUW - 95 - 23\\
\medskip
\hfill  PITHA - 95/24\\
\medskip
\hfill  gr-qc/9511081\\
\medskip

\begin{center}
{\LARGE {Classical and Quantum  Gravity in 1+1 Dimensions\\
Part II: The Universal Coverings}
 \\ \medskip  {}}
\medskip
\vfill
            
\renewcommand{\baselinestretch}{1}
{\large {THOMAS
KL\"OSCH\footnote{e-mail: kloesch@tph.tuwien.ac.at} \\
\medskip
Institut f\"ur Theoretische Physik \\
Technische Universit\"at Wien\\
Wiedner Hauptstr. 8-10, A-1040 Vienna\\
Austria\\
\medskip
\medskip THOMAS STROBL\footnote{e-mail:
tstrobl@physik.rwth-aachen.de} \\ \medskip%\medskip
%\medskip \medskip
Institut f\"ur Theoretische Physik \\
RWTH-Aachen\\
Sommerfeldstr. 26-28, D52056 Aachen\\
Germany\\} }
\end{center}

\setcounter{footnote}{0}
\renewcommand{\baselinestretch}{1}                          %!!!

\begin{abstract}

  A set of simple rules for constructing the
  maximal (e.g.\ analytic) extensions for any metric with a Killing
  field in an (effectively) two-dimensional spacetime is formulated.
  The application
  of these rules is extremely straightforward, as is demonstrated at
  various examples and illustrated with numerous figures. 
  Despite the resulting simplicity we also comment
  on some subtleties concerning the concept of Penrose diagrams. Most
  noteworthy among these, maybe, is that (smooth) spacetimes which
  have both degenerate and non-degenerate (Killing) horizons do not
  allow for globally smooth Penrose diagrams. Physically speaking this
  obstruction corresponds to an infinite relative red/blueshift between
  observers moving across the two horizons. -- The present work
  provides a further step in the classification of all global
  solutions of the general class of two-dimensional
  gravity-Yang-Mills systems introduced in Part I \cite{partI},
  comprising, e.g., all generalized (linear and nonlinear) dilaton
  theories. In Part I we constructed the local solutions, which were
  found to always have a Killing field; in this paper we provide all
  universal covering solutions (the simply connected maximally
  extended spacetimes). A subsequent Part III \cite{partX} will
  treat the diffeomorphism inequivalent solutions for all
  other spacetime topologies.

  Part II is kept entirely self-contained; a prior reading of
  Part I is not necessary. 

\end{abstract}

%\vfill \noindent%
%PACS-Numbers:
%04.20.Dw % Singularities and cosmic censorship
%04.20.Gz % Spacetime topology, causal structure, spinor structure
%04.60.Kz % Lower dimensional models; minisuperspace models
%04.70.Bw % Classical black holes

\vfill
\noindent to appear in {\em Class.\ Quantum Grav.} \hfill November 1995  \\  
\end{titlepage}

\renewcommand{\baselinestretch}{1}
\small\normalsize

\section{Introduction}

\plabel{Intro}

As is well known, in two spacetime dimensions any metric is
conformally flat: Using null coordinates $u$ and $v$, $g$ takes the
form (conformal gauge)
\be g=\exp[\rho(u,v)] \, du dv \, .
  \label{confo} \ee
So, at first sight it might appear that the causal
structure of any 2d metric is the same as the one of a flat metric.
Certainly this is far from true, as is seen from the various models of
dilaton gravity or already from spherically symmetric 4d gravity.
In many cases the nontrivial global structure comes about, because the
(maximal) domain of definition of the function $\rho$ in \re{confo})
is only a subset of $\dR^2$. For the (spherically reduced) Schwarzschild
solution this is a well-known feature of the Kruskal coordinates; for
Reissner-Nordstr\"om, or more complicated 2d metrics, the same
mechanism applies in a less trivial way.
Sometimes, moreover, it is not possible to stay within one plane sheet;
the conformal gauge can only be achieved by allowing overlapping layers
(cf.\ Fig.\ 6). And even within one sheet smoothness considerations
may obstruct a global attainability of \re{confo}) (cf.\ Fig.\ 3).
Thus, the global causal structure of general 1+1 metrics may be quite
involved, even if, as in the present paper, attention is restricted to
1+1 metrics with Killing vectors.      

Somewhat less popular than the conformal gauge \re{confo}), but for some
purposes better suited, is the chiral gauge 
\be g=2dx^0dx^1 + k(x^0,x^1) \,
  dx^1 dx^1 \, . \label{lcgauge} \el chiral
It results upon the choice of $x^1$ as a
label for one set of null-lines and of $x^0$ as an appropriately normalized
affine parameter on them; the normalization (or synchronization) of the affine
parameters may be prescribed on any line $x^0 = c=const$ by requiring that
there the ``speeds'' $\6_0|_{c,x^1}$ have unit inner product with $\6_1$.
Equivalently: Choose $x^1$ as above and subsequently, with these
``provisional'' coordinates, $x^0$ as
$\int g_{\tilde 01}(\tilde x^0, x^1) d \tilde x^0$.

If $k$ in \re{lcgauge}) does not depend on $x^1$, this metric has a Killing
field $\6_1$.  But, remarkably, also the converse is true: {\em Any\/} 2d
metric with a Killing vector allows for coordinates such that locally it
takes the generalized Eddington-Finkelstein form
\be
  g=2dx^0dx^1+h(x^0)(dx^1)^2 \, .  \el 011h 
It is obtained by labelling
Killing trajectories and (locally) transversal null-lines by $x^0$ and
$x^1$, respectively, where $x^1$ is the Killing-flow parameter and $x^0$ an
affine parameter of the chosen null-lines normalized as before. To arrive at
\re{011h}) one can repeat the construction of \re{chiral}), but now choosing
$x^0 = c$ to be a Killing trajectory and $x^1$ a Killing-flow parameter along
this line.
For a given metric $g$ the function $h$ in \re{011h}) is generically unique
up to an equivalence relation $h(x^0) \sim {1\0{a^2}}h(ax^0+b)$, $a,b=const$.
%(since $x^0$ has to be an affine parameter for the null-extremals).
Only for Minkowski and deSitter space this is not quite true, because they
have more than one Killing field (in fact three). Minkowski space can be
described by $h=ax^0+b$ and (anti-)deSitter space of curvature $R$ by
$h={R\0 2}(x^0)^2+ax^0+b$ (since $R\equiv h''$).% FOOTNOTE
\footnote{Different choices of $h(x^0)$ correspond to different Killing
  fields $\6_1$. For instance, in the case of Minkowski space $h=const$
  implies that $\6_1$ generates translations (timelike, null, or spacelike,
  according to $\mbox{sgn}\,h$), whereas $h$ linear in $x^0$  ($a\ne0$)
  corresponds to boosts.
  For deSitter cf.\ the paragraph around \re{deSitter}) and Fig.~9.}

Our motivation for studying 2d spacetimes with a Killing field is that
metrics of such a type arise in the analysis of the very general class of
2d gravity-Yang-Mills systems introduced in Part I \cite{partI} (cf.\ also
Sec.\ 2): All the solutions of, e.g., generalized (linear or nonlinear)
dilaton theory have a Killing field. This continues to hold for
generalizations with nontrivial torsion as well as for a dynamical
(possibly dilaton dependent) coupling of these theories to a Yang-Mills
field of an arbitrary gauge group. In all of these cases we brought the
metric $g$ into the form \re{011h}) and determined the possible functions
$h$. For any fixed pure gravity theory there is a one parameter family of
such functions $h$ (generalized Birkhoff theorem) while in the presence of
a Yang-Mills field a second parameter arises. In the space of all possible
Lagrangians considered in Part I, furthermore, any function $h$ can be
generated.

Clearly, in general a coordinate system \re{011h}) is attainable locally
only. For instance, the Killing trajectories $x^0=const$ may become parallel
to the null-lines $x^1=const$ somewhere or they may run into a point where
the Killing vector field vanishes.
In contrast to the more general charts \re{lcgauge}), by construction
\re{011h}) is applicable only in the neighbourhood of points where the
Killing vector does not vanish. (We constructed a chart valid around points
of vanishing Killing vector, too, cf.\ Eq.\ \re{saddle}) below; but again,
in general also these charts are local only).  For physical reasons, however,
one is interested in maximally extended spacetimes: All extremals (which, if
timelike, describe the motion of test particles) should be either complete
(have infinite length) or run into a true curvature singularity. 

Thus, in the present paper, we will study the maximal extension of the
local charts \re{011h}). Certainly, in general such an extension is not
unique. Already the (analytic) Minkowski solution, $h \equiv 0$, allows
for inextendible manifolds of planar, cylindrical, or toroidal topology
(the latter two even carrying further continuous parameters). 
The Minkowski-plane, however, is the unique universal covering space of
the cylinders and tori and they, vice-versa, are obtained by factoring out
a discrete symmetry group from this covering manifold. This generalizes:
Any multiply connected solution, however complicated it may be, can be
obtained from the universal coverings (i.e. the maximally extended {\em
  simply connected\/} manifolds) by factoring out a discrete symmetry group
(cf., e.g., \cite{Wolf}). Therefore, in the present paper we will restrict
ourselves to the simply connected extensions. The global solutions with
non-trivial spacetime topology will be dealt with then in the following
paper \cite{partX}. 

Still even if we restrict ourselves to simply connected spacetimes the
extension of a local solution of the form \re{011h}) is not unique without
additional specifications. Take, e.g., the case that the function $h$ is
given in the interval $[0,1]$ and known to vanish there. Obviously already
within a chart of the type \re{011h}) there is an infinity of smooth
extensions of the function $h$ to values of $x^0$ outside the given
interval. But even if the function $h$ is known on a maximal domain of
$x^0$, having a boundary only if there $h'' \equiv R \to \pm \infty$, there
are similar ambiguities in a smooth extension of the (generically) local
chart \re{011h}) into regions not covered by that chart. One way to avoid
such ambiguities is to require that $g$ should be analytic everywhere.
Alternatively, working in the framework of a given gravity-Yang-Mills model
considered in Part I, we may require the field equations of the given
Lagrangian to restrict the extension. In this case the universal covering
solutions turn out to be unique also for non-analytic functions $h$.
Either of these two philosophies underlies the simple extension rules
derived in the present paper. 

\vskip3mm

The organization of this paper is as follows: In Sec.\ \ref{Review} we
recollect some of the results of Part I. This section serves mainly to
recall the generality of models treated by us as well as to provide some
explicit examples used later on for illustration. It may well be omitted on
a first reading, if one is not so much interested in 1+1 gravity models but
merely in the, say analytic, extension of a metric \re{011h}). In Sec.\
\ref{General}, then, we discuss some general properties of the metric
\re{011h}) and use its symmetries to solve the equation of extremals.

Sections \ref{Blocks} and \ref{Saddle} are the heart of the present
paper: Here we derive a simple building block principle for obtaining the
universal covering solutions of any metric with local form \re{011h}). In
the first of these two sections the charts underlying \re{011h}) are mapped
into a finite subdomain of $\dR^2$ with null-lines as Cartesian
coordinates. These are the basic building blocks for Penrose diagrams,
which are glued together in Section \ref{Saddle}. The considerations are
described by means of an illustration for a typical function $h$, cf.\ 
Figs.\ 2 and 5.  Our method is a streamlined and generalized version of the
one suggested previously by M.\ Walker \cite{Walker} (cf.\ also \cite{Brill})
for a rather restricted class of metrics. Also we point out some possible
pitfalls within this method (ignored for the most part in the
previous literature). The most severe one is that given a function $h$ with
zeros of different orders, despite the existence of a smooth inextendible
covering space there is, in this case, {\em no\/} globally smooth Penrose
diagram for it. The physical reason for such an obstruction is found in an
infinite red/blue-shift (Fig.\ 3). Another issue concerns the boundary lines 
of Penrose diagrams. While any single space- or time-like boundary line may
be straightened by a conformal diffeomorphism, in some cases it is again
global smoothness that forbids to perform this operation to all boundary
lines of a Penrose diagram at the same time, thus leaving some of these
boundary lines bent out- or inwards (Figs. 4,7,11).

The resulting method to construct Penrose diagrams for a given metric
\re{011h}) is summarized in terms of simple rules in Sec.\ \ref{Recipe}.
The list of rules is followed by their
application to the three examples provided at the end of Sec.\ 
\ref{Review}. This demonstrates the efficiency of the method. In its final
form one merely has to determine the number and degrees of the zeros of the
function $h$ as well as its asymptotic behaviour. The rest is simple
pictorial gluing.

On the level of purely local considerations the models considered in Part I
show many similarities. However, taking global aspects into account, there
arise relevant differences. This becomes particularly pronounced, when
solutions on non-simply connected spacetimes are constructed
(cf.\ Part III \cite{partX}). We give a first flavour of this in the
concluding Sec.\ \ref{Outlook} and comment on related issues relevant for the
quantum theory (Part IV \cite{partX}). 

\section{Recollection of Results of Part I}

\plabel{Review}

The model we deal with reads in its most general form
\be L =
  \int_M \,\left[X_a De^a + X^3 d\o + W(X_aX^a,X^3) \, \varepsilon
  \right] \,\,+\,\, \int_M \, Z(X_aX^a,X^3) \, \trace (F \wedge \ast
  F) \,\, , \el{grav}
consisting of a gravity- and a Yang-Mills part. Here
$e^a$ with $a \in \{+,-\}$ is the zweibein in a light cone basis of the
frame bundle, $\varepsilon \equiv e^- \wedge e^+$ is the corresponding
volume form ($\Rightarrow \varepsilon^{+-}= +1$), and $\o$ (or $\o^a{}_b
\equiv \varepsilon^a{}_b \o$) is the Lorentz or spin connection; so
$d\o$ is the gravity curvature two-form and $De^a \equiv de^a +
\varepsilon^a{}_b \o \wedge e^b$ the torsion two-form.  $F$ is
the curvature two-form of the Yang-Mills gauge fields, the trace
represents some non-degenerate inner product on the Lie algebra, and the
Hodge dual operation ``$\ast$'' is performed by means of the volume form
$\varepsilon$. $X^a$ and  $X^3$  are dilaton-like
dynamical fields, which in the Hamiltonian formulation will serve as
momenta to the one-components of $e^a$, $\o$, respectively (cf.\ Part IV
\cite{partX}). $W$ and $Z$ are some arbitrary potentials (with $Z
\neq 0$ or $Z \equiv 0$) of $X^3$ and the Lorentz-invariant combination 
$X_aX^a \equiv 2 X^+X^-$.  In the formulation above the metric $g$ is
obtained via $g = 2 e^-e^+ \equiv e^- \otimes e^+ + e^+ \otimes e^-$.

The action \re{grav}) allows also to treat generalized 2d dilaton gravity
\cite{Banks} coupled to Yang-Mills fields:
\be L^{gdilYM}=\int_M d^2x
  \sqrt{-\det g} \,\, \left[D(\Phi) R + \mbox{$\2$} g^{\m\n} \6_\m 
  \Phi\6_\n \Phi - U(\Phi) + B(\Phi) \, \trace F_{\m\n} F^{\m\n}\right] \, , 
  \el gdil
where $R$ is the Ricci scalar of $g$ and $\Phi$ is the dilaton field.
In order to describe \re{gdil}) by means of the general action \re{grav}),
however, $e^a$ is no longer the standard zweibein; rather, now
\be g= K(X^3)
  \, e^+ e^- \el gconf
where the conformal factor $K$ is determined by the
potential $D$ in \re{gdil}) (for further details cf.\ Part I). 
In this case $X^3=D(\Phi)$, furthermore, and $W$ and $Z$
in \re{grav}) are $X^a$-independent so that the $X^a$-fields become
Lagrange multiplier fields enforcing torsion zero (this holds also despite
the complication \re{gconf})). We note on this occasion that for the
same reasons as those given in the introduction the transformation
\re{gconf}), although conformal, has in some cases implications on the global
causal structure of $g$ (cf.\ related remarks in Part I as well as
\cite{KKL}). In contrast to minimally coupled massless scalar fields,
furthermore, the Yang-Mills fields ``feel'' the conformal factor of the metric.

Let us make a side-remark: Obviously, in the transition from \re{gdil}) to
\re{grav}) one of the three potentials went into the redefinition of field
variables (into \re{gconf}) and/or $X^3=D(\Phi)$).  As shown implicitly in
Part I, further field redefinitions allow to eliminate also these remaining
two potentials in \re{grav}), reducing the action effectively to the one
for a set of free abelian gauge fields. (Just implement the transformation
to Casimir-Darboux coordinates, $X^a,X^3 \to \mytilde X^i$ etc., on the level
of the Lagrangian, cf., e.g., \cite{Heiko}). Such a procedure is
legitimate, if one is interested in {\em local\/} solutions only, which, as
in the present paper, are then extended to global ones by patching. This
allows us to drop also the restriction $D'\neq 0$, made in Part I. On any
of the branches of $D$ with $D' \neq 0$ the transition from \re{gdil}) to
\re{grav}) is legitimate and the latter Lagrangian (or the further
simplified version mentioned above) describes the corresponding local
solutions perfectly well.  However, when discussing the quantum theory of
\re{gdil}) via the formulation \re{grav}) in Part IV, the restriction $D'
\neq 0$ (everywhere) may be essential.

\medskip

In Part I it was shown that for the general class of models described
by \re{grav}) or \re{gdil}) nearly everywhere the local solution for
$g$ may be put into the form \re{011h}) with $h(x^0)$ as specified
there. E.g.\ for \re{grav}) without Yang-Mills fields (i.e.\ 
$Z(X_aX^a,X^3)\equiv0$) and without torsion (i.e.\ $W(X_aX^a,X^3)\equiv
V(X^3)/2$), we found
\be h(x^0)=\int^{x^0}V(z)dz+C \, ,
  \label{Potential} \ee
where the ``Casimir constant'' $C$ is a meaningful
constant of integration (related to the total mass). While in this case the 
free parameter $C$ merely shifts $h$ ``vertically'', the dependence on this
parameter becomes more involved for theories with torsion or in the context
of \re{gdil}, \ref{gconf}). In the presence of Yang-Mills fields 
$h$ depends on a second parameter, the ``charge'' $q= Z^2 \, \trace (\ast 
F)^2=const$. Let us note (cf.\ Part I), furthermore,  that the 
gravitational sector of any gravity-Yang-Mills system \re{grav}) may be
described equivalently through \re{grav}) with  potentials 
\be \mytilde W:= W + {q \0 Z} \; , \quad \mytilde Z := 0 \, 
  \label{poteff} ,\ee
now to be analyzed for all possible values of $q$. Thus in particular 
a minimally coupled ($Z \equiv const.$)  gauge field 
contributes to the metric $g$ merely via a dynamical cosmological
constant. Taking all the Yang-Mills potentials into account, on the other
hand, the local solutions have been found to be parametrized by $r-1$
further constants besides $C$ and $q$, where $r$ denotes the rank of the
Lie algebra.

In the chiral gauge \re{011h}) also the other fields depend only on the
coordinate $x^0$ so that $\6_1$ generates symmetries of the full solutions
to the field equations.  In the case of
\re{Potential}), e.g., $X^3=x^0$ and $X_aX^a=h(x^0)$. Generally,
the coordinate $x^0$ in \re{011h}) always either ranges over
all of $\dR$ or, if not, then on the boundary $X^3$ (or some
``back-transformed'' dilaton field) diverges.% FOOTNOTE
\footnote{In the case that $D$ has extrema, a prior gluing of solutions
  corresponding to successive branches may be necessary.}

The chart \re{011h}) only fails at zeros of the Killing field.  For
isolated simple zeros (say, at $x^0=a$) we have already in Part I
(\cite{partI}, Eqs. (64,65)) obtained the chart
\be g=-{4\0{h'(a)}}\left[dxdy+{xy-{h(xy+a)\0h'(a)}\0x^2}dx^2\right] \, ,
  \el saddle
$h$ being the same function as before (but cf.\ also the
alternative approach of Sec.\ \ref{Saddle}, Eqs.\
\re{Krtrfu}--\ref{Kruskal})).
Note that the chart \re{saddle}) is, after a trivial rescaling, still of the
form \re{lcgauge}), where, however, now the lines of constant $y \propto x^0$
are different from the Killing trajectories and $x \propto x^1$ does not 
coincide with a Killing flow parameter. Consequently, in contrast to
\re{011h}) the form \re{saddle}) allows to cover also simple zeros of the
Killing field. Higher order zeros, on the other hand, will turn out to
lie always in infinite distance, so they cannot be
covered by a smooth coordinate system.% FOOTNOTE
\footnote{Alternatively to this one may argue equally well that the
  coordinates in the {\em target} space (of the $\s$-model formulation of
  Part I) corresponding to the charts \re{011h},\ref{saddle}) provide an
  atlas of the respective symplectic leaf. The absence of a chart of the type
  (\ref{saddle}) for zeros of $h$ of higher degree is then explained by the
  fact that in the target space such points constitute independent (pointlike)
  symplectic leaves, thus giving rise to other independent
  solutions, namely the  deSitter solutions \re{deSitter}).}
Since in two dimensions a Killing
field vanishing along a line must vanish everywhere, this exhausts all
cases.

Finally, if $W(0,X^3)$ in \re{grav}) has zeros (we call the corresponding
values of $X^3$ ``critical'', $X^3_{crit}$), then one gets additionally the
``deSitter'' solutions $X^a\equiv 0, X^3 \equiv X^3_{crit}$,
\be De^a =0 \, ,\,\, d\omega =
  -{\partial W \0 \partial X^3}(0,X^3_{crit})\e \,, \el deSitter
which have vanishing torsion and constant curvature all over $M$.%
% FOOTNOTE
\footnote{According to \re{gconf}) these solutions remain to be
  ``deSitter'' also in the context of \re{gdil}). If $D$ has extrema,
  furthermore, then in some cases there will also arise solutions of
  constant $\Phi$ corresponding to the critical values of $D$.} 
The metric for such a solution can be brought into the form \re{011h}),
too, with $h(x^0)={\partial W \0 \partial X^3}(0,X^3_{crit}) \cdot
[(x^0)^2 +d], d=const$.  As remarked in the introduction,  the
coordinate system underlying \re{011h}) is adapted to a Killing direction. In
the case of deSitter space the Killing fields form a {\em
three\/}-dimensional vector space, in which the vector fields $\6_1$
corresponding to the different choices  $d=-1,0,1$ in $h$ are pair-wise
independent (for a basis take, e.g., two of them and their Lie bracket).%
% FOOTNOTE
\footnote{As we will find below, the universal covering solutions of 2d
  deSitter or anti-deSitter space may be described by a ribbon-like conformal
  diagram.  Fig.\ 9 displays such Penrose diagrams for deSitter space where
  the Killing lines $x^0 = const$ have been drawn; {\bf J1} corresponds to
  $d<0$, {\bf J2} to $d=0$, and {\bf J3} to $d>0$.}

\medskip

We conclude this brief review with three specific models comprised in
the above formalism (cf.\ Part I) which
will serve as examples in Sec.\ \ref{Recipe}.  First the so-called
Jackiw-Teitelboim (JT) model of 1+1 (anti)deSitter gravity \pcite{JT} 
obtained by the choice $2W := V(X^3) = \L X^3$ in \re{grav}):
\be L^{JT}=-\frac12\int_M d^2x\sqrt{-\det g} \; X^3 \; (R-\L) \quad
  \Rightarrow \quad h^{JT}(x^0)=C-{\L\02}(x^0)^2 \, ; \el JT
here $X^3$
serves as Lagrange multiplier to enforce the field equation $R = \L \equiv
const$, while the $X^a$-fields were used in \re{grav}) to ensure torsion
zero. In contrast to the deSitter solutions \re{deSitter}) present also in
most of the generic models \re{grav}), here the fields $X^i$ are not
constant all over $M$ (in particular, e.g.,  $X^3=x^0$). Consequently in
the case of the $JT$-model the fields $X^i$ break the isometry group of
the deSitter metric down to a one-dimensional group; therefore, in contrast
to $d$ in the function $h$ of \re{deSitter}), here the constant $C$
distinguishes diffeomorphism inequivalent solutions of the field
equations.% FOOTNOTE
\footnote{Note, however, that also for the JT-model there are the 
  ``critical'' solutions $X^a \equiv 0, \, X^3 \equiv 0$ of the
  symmetry-unbroken type \re{deSitter}).}

The potential $W^{R^2}:= -(X^3)^2 +\L$, as the second example, leads, upon
elimination of the $X$-coordinates, to the Lagrangian of two-dimensional
$R^2$-gravity,
\be L^{R^2}= \int_M d^2x
  \sqrt{-\det g} \, (R^2/16 + \L) \quad \Rightarrow \quad
  h^{R^2}(x^0)=-\frac23(x^0)^3+2\L x^0+C \, . \el R2
As an example for non-trivial torsion, the potential $W^{KV}= - \a X_aX^a/2
- (X^3)^2 +\L/\a^2$ allows to describe the
Katanaev-Volovich model \pcite{KV}
\ba L^{KV} &=& \int [-{1 \0 4} d\o \w
  \ast d\o - {1\0 2\a} De^a \w \ast De_a + {\L \0 \a^2} \e] \quad \Rightarrow
  \nonumber\\ h^{KV}(x^0) &=& {1 \0 \a} \left\{ C x^0 - 2 (x^0)^2 [(\ln
  x^0-1)^2+1-\Lambda] \right\} \, .\la{KV} \ea
Here $x^0$ ranges  over
$\dR^+$ only. In the latter two models $X^3$ became proportional to 
the curvature scalar $R$ on shell, while $X^a$ is a Lagrange multiplier for
torsion zero in \re{R2}) and proportional to torsion $\ast De^a$ in
\re{KV}), respectively.

For more details on these and other models consult Part I \cite{partI} and
the references given there.

\section{General Remarks and Extremals}

\plabel{General}

To find the maximal extension of the local solutions \re{011h}),  we want to
exploit the symmetries of this metric. As pointed out already in the
introduction \re{011h}) allows for a Killing field, namely
$\partial\0\partial x^1$. The function $h(x^0)$ measures the norm squared
of this Killing vector, furthermore. In particular a zero of $h$ indicates
that the line $x^0=const$ is a light-like (null) Killing trajectory. We
will adopt the customary term {\em Killing horizon\/} for such lines and
{\em degenerate Killing horizon\/} if the corresponding zero of $h$ is of
higher order (i.e., $h'|_{h=0}=0$). Also we will call the regions between
two zeros (or beyond the first/last zero) {\em sectors\/}. Finally, we call
a sector {\em stationary\/} if the Killing field is timelike there ($h>0$)
and {\em homogeneous\/} if it is spacelike ($h<0$).

The metric \re{011h}) allows also for another symmetry, seen best in
coordinates
\be r := x^0 \, , \;\; t := x^1 + f(x^0)
  \el rtKoord
with
\be f(x^0) \equiv \int^{x^0} {du \0 h(u)} +\hbox{const} \, .
  \el fun
These
coordinates are well-defined wherever $h \neq 0$ and bring the metric \rz
011h into the generalized Schwarzschild form
\be g= h(r) dt^2 -{1 \0 h(r)}
  dr^2 \, , \el gSS
where $h$ is the {\em same} function as in \re{011h}).
\rz gSS has the disadvantage of becoming singular at zeros of $h$, whereas
\re{011h}) behaves perfectly well there. However, it displays a further
independent (discrete) symmetry, namely one under an inversion of the
Killing parameter $t$, $t \leftrightarrow -t$.  If $h>0$, such that the
transformation amounts to a time-reversal, then the metric is called {\em
  static\/}, and thus we see that in two dimensions stationary implies
static. This comes, however, as no surprise since in two dimensions any
vector field is hypersurface-orthogonal.% FOOTNOTE
\footnote{A static metric can also
  be characterized by the condition that the timelike Killing field be
  hypersurface-orthogonal.} In connection with the coordinates \re{011h})
this time- (or space-)reversal symmetry will prove to be a powerful tool in
constructing the maximal extension.  Therefore we write it down explicitly
also in the original $x^0,x^1$-coordinates:
\be {\mytilde x^0} =x^0 \quad , \qquad
  \mytilde x^1= -x^1-2f(x^0) \el glue
with $f$ as in \re{fun}). Of course, this symmetry works only within one
sector at a time, because on
its edges (zero of $h$) $f$ diverges.  We will call this transformation
{\em flip}.  Note that changing the constant in $f$ yields a different
transformation (shifting the origin of the Killing-parameter $t$ and thus
the reflexion axis), so actually one has a one-parameter family of
flip-transformations.
The effect of this flip transformation on the dynamical fields is exactly
that of the gauge changing transformation Eq.\ (61) in Part I. Thus the flip
is in fact a symmetry transformation for the full solution to the models
\re{grav}) or \re{gdil}).

\vspace{.3cm}

The standard method for constructing the maximal extension involves
the determination of the extremals 
and the study of their completeness properties. 
Since we will directly apply the 
transformation \re{glue}) to extend our local solutions, 
 this is not really  necessary in the present case.
Still let us solve their equation  
for illustrative purposes; after all, extremals, i.e.\ curves extremizing 
the action $m\!\int\!ds^2$, are supposed to describe the motion of a test
particle in the curved background and are used to study the completeness
properties of the solutions.

Using the Christoffel symbols 
$\G_{\m\n\r} \equiv (g_{\m\n,\r} + g_{\r\m,\n} - g_{\n\r,\m})/2 \,$, 
the equation for the extremals reads 
\be \ddot x^\m  +  \G^\m{}_{\n\r} \dot x^\n \dot x^\r =0\,.
  \el extrem1
For the metric  \re{011h}) this reduces to 
\be \ddot x^0=-h'\dot x^1 (\dot x^0+ {h\over2} \dot x^1)
  \quad, \qquad
  \ddot x^1 =   {h'\over2}(\dot x^1)^2 \, , \el extrem2
with $h'\equiv dh(x^0)/ dx^0$. 
Up to affine transformations, Eq.\ \re{extrem1}) yields a unique
parametrization of the solutions $x(\t)$. For non-null extremals 
the arclength $s$ itself is such an {\em affine parameter\/}. 
Let us remark that in the torsion-free case extremals
coincide with autoparallels, satisfying $\nabla_{\dot x} \dot x = 0$, while 
in the case of nonvanishing torsion they are autoparallel only
with respect to the Levi-Civita connection.

Clearly in two dimensions any null-line $g(\dot x, \dot x)=0$
solves \re{extrem1},\ref{extrem2}). In the chart \rp 011h
these {\it null extremals} are:
\ba   x^1&=&\hbox{const} \, , \pl{null1} \\
{dx^1\over dx^0}&=&-{2\over h} \quad,\quad \mbox{wherever }h(x^0) \neq 0,
  \pl{null2a} \\
 x^0&=&\hbox{const} \, ,\quad \mbox{if} \,\, h(x^0)=0  \, .\pl{null2b} \ea
The extremals \re{null2b}) are exactly the Killing horizons mentioned above.
Plugging these solutions back into (\ref{extrem2}), we see that for
\re{null1}) and \re{null2a}) the affine parameter is $\t=ax^0+b$,
($a,b=const$).
Thus these null extremals are complete at the boundaries, iff the coordinate
$x^0$ extends to infinity into both directions of the charts \re{011h}). For
the models under study this is the case e.g.\ for all torsionless theories,
whereas the KV-model \re{KV}) provides an example 
of incomplete extremals \re{null1}, \ref{null2a})   at $x^0=0$ (which  
is still a true singularity as $R \propto \a X^3 = \ln x^0/\a^2$,
\cite{partI}, blows up there).
The affine parameter for the extremals \re{null2b}) depends on the kind
of zero of $h(x^0)$: For non-degenerate horizons ($h'(x^0)\ne0$) we get
$\t=a\exp(-{h'\0 2}x^1)+b$, so they are incomplete on one side, whereas
degenerate horizons ($h'(x^0)=0$) are always complete, as then
$\t=ax^1+b$.
 
The non-null extremals are found most easily by making use of the constant 
of motion associated with the Killing field $\partial\over\partial x^1$ 
(cf., e.g., \pcite{Thi}): $g({\partial\over\partial x^1},\dot x)
\equiv \dot x^0+h\dot x^1=const$ along extremals. Knowing that for non-null 
extremals we may choose the length as an affine parameter, 
this equation may be rewritten in the form 
\be (dx^0+hdx^1)=\hbox{const}\cdot ds = \hbox{const} \cdot
  \sqrt{\Big|2dx^0dx^1+h(dx^1)^2\Big|} \,. \ee
The resulting  quadratic equation has the solutions
\ba {dx^1\over dx^0}&=&{-1\pm\sqrt{\displaystyle{c\over c-h(x^0)}}\over
  h(x^0)} \quad,\quad c=\mbox{const}\, , \pl{exsol1} \\
  \noalign{\vspace{-0.3 cm}}\nonumber\\
  x^0&=&\hbox{const},\quad \mbox{if} \,\, h'(x^0)=0 \, . \pl{exsol2} \ea
\re{exsol1}) is meant to hold  only  when  meaningful; the
condition in \re{exsol2}) is immediate from \re{extrem2}).
The null-extremals \re{null1},\ref{null2a}) can be obtained from
\re{exsol1}) as the limiting case $c\rightarrow\infty$.
Eq.\ \re{exsol1}), even if it cannot be integrated explicitly, allows for 
a comprehensive
qualitative discussion of the extremals. For brevity we do not go into
details here.  Let us just point out that under the flip transformation
\re{glue}) each extremal \re{exsol1}) is mapped onto another one with the same
value of $c$, but with the opposite sign of the square root. Similarly the
two types of null extremals \re{null1}) and \re{null2a}) get interchanged.
It is straightforward to see that the line element for the extremals
\rp exsol1 is
\be ds={1\over\sqrt{|c-h(x^0)|}}dx^0 \, ,
  \el 11
which will be used to determine their completeness properties. 
The extremals \re{exsol2}), on the other hand, are obviously always complete, 
since $x^1 \propto s + const$ ranges over all of $\dR$.

\section{Construction of the Building Blocks, Penrose Diagrams}

\plabel{Blocks}

In this and the following section we will 
derive the general rules of how
to find the Penrose diagrams starting from any given metric of the form
\re{011h}). In section \ref{Recipe} then we will summarize the resulting 
 simple building block principle and  apply it to
the specific models (\ref{JT}, \ref{R2}, \ref{KV}) for illustration.

Let us first assume that $h$ has no zeros. Then the diffeomorphism
\be x^+=x^1 + 2f(x^0),\quad x^- =x^1
  \el conf
with $f(x^0)$ as before \re{fun}) 
brings the solution into conformally flat form, i.e., 
the metric in the new coordinates reads $g=h(x^0)dx^+dx^-$.
The Killing field $\partial\over\partial x^1$ then becomes
${\partial\over\partial x^+}+{\partial\over\partial x^-}$ and the flip
transformations \re{glue}) are simply the reflexions at any of the
lines $x^++x^-=const$.

Now, according to the asymptotic behaviour of $h$ there are three cases      
to be distinguished: First, the range of $f$ may be all of $\dR$
(e.g. for $h\sim{(x^0)}^{k\leq1}\hbox{\ as $x^0 \to \pm\infty$}$). Then the
image of the diffeomorphism \re{conf}) is the whole $(x^+,x^-)$-coordinate
plane. Second, the range of $f$ may be bounded from one side (e.g.\ for
$h\sim{(x^0)}^{k>1}\hbox{\ as $x^0 \to \pm\infty$}$). In this case the image
is only the half plane% FOOTNOTE
\footnote{This reflects the fact that then in the original
  $(x^0,x^1)$-coordinates any of the null extremals \re{null2a}) has one of
  type \re{null1}), i.e.\ $x^1=const$, as an asymptotic and thus the 
  null-lines \re{null2a}) do not intersect all of the null-lines \re{null1}).}
bounded by the line $x^+-x^-=\lim f$.
Finally, $f$ may be bounded from both sides, in which case the image is a
ribbon $a<x^+-x^-<b$.

\begin{figure}[h,t]
\begin{center}
\leavevmode
\epsfxsize 8cm \epsfbox{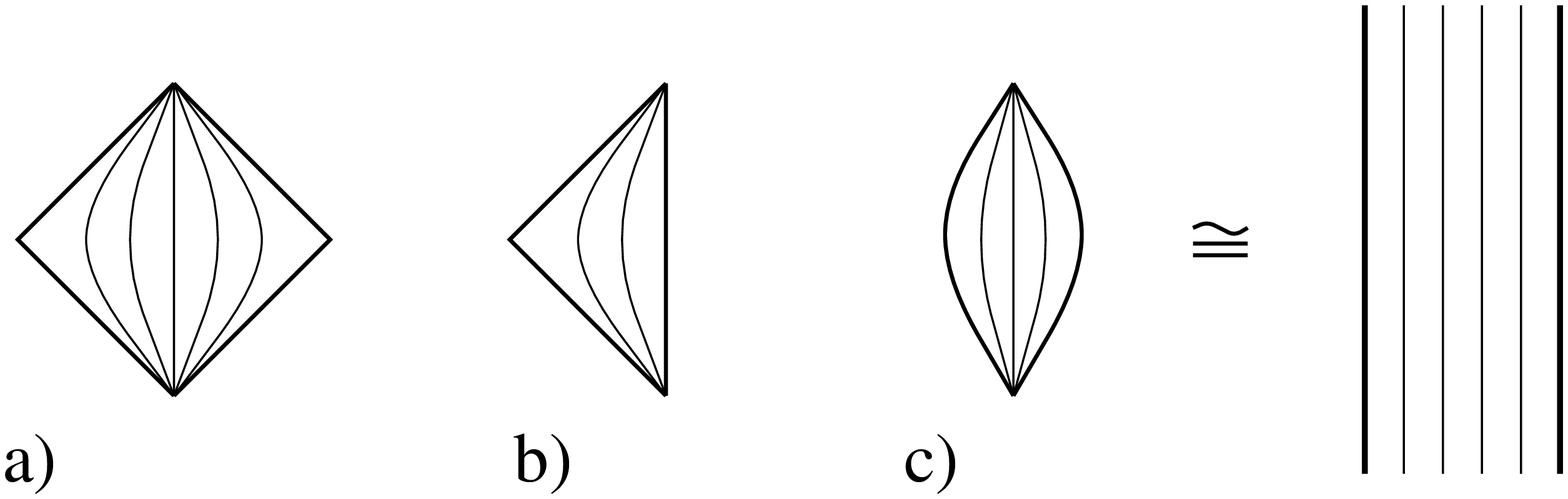}
\end{center}
\renewcommand{\baselinestretch}{.9}
\small \normalsize
{\bf Figure 1:} {\small Penrose diagrams for $h>0$ (thin lines represent
Killing-trajectories).}
\end{figure}
In order to get the Penrose diagrams we further apply a conformal
diffeomorphism like $x^\pm\rightarrow\tan x^\pm$ which maps the solutions into
a finite region, and finally we turn the patch 45 degrees (counter-)clockwise
for $h\,(<)\!>0$,
given our convention that a positive (negative) $ds^2$ corresponds to a
timelike (spacelike) distance.% FOOTNOTE
\footnote{Remember that $h$ measures the norm squared of the Killing field.
  Thus $h\!<\!(>)\,\,0$ corresponds to homogeneous (stationary) spacetimes.}
The three cases then correspond to a square, triangle or a lens-shaped region,
resp.\ (cf.\ Fig.\ 1; the diagrams for $h<0$ are similar but turned by 90
degrees). In the last case, however, we will sometimes prefer the
(uncompressed) infinite ribbon form.

\medskip

Let us now come to the cases where $h$ has zeros.
These shall be treated by means of a generic example, the
function $h$ of which is drawn in Fig.\ 2a.
In Fig.\ 2b we qualitatively depicted  representatives of the null extremals
\re{null1}, \ref{null2a}, \ref{null2b}) corresponding to the metric of
Fig.\ 2a.
Now, within each sector the diffeomorphism \re{conf}) could be applied (again
exhibiting the flip symmetry of the sector);
it, however, breaks down at $h(x^0)=0$.
It may be cumbersome to write down explicitly the diffeomorphism that
brings $g$ into conformal form on all of the chart underlying Fig.\ 2b.

Fortunately, the explicit form of such a diffeomorphism need not be
constructed; we can proceed with a simple geometric argumentation:
By an $x^1$-dependent distortion of the $x^0$-coordinate one can
straighten the null extremals \re{null2a}), leaving the horizons
\re{null2b})  as well as the null extremals \re{null1}),
i.e.\ $x^1=x^- = const$, unmodified.
Note that in our example the $(x^+, x^-)$-chart
cannot be all of $\dR^2$ any more; rather on the right-hand side there will be
some boundary, because due to the asymptotic behaviour of $h$ the null lines
of type \rp null2a do not intersect all
null lines $x^- = const$. By means of a subsequent conformal diffeomorphism
(like $x^- \to \tan x^-$; cf., however, the following paragraph)
and a similar one for $x^+$
the new coordinate chart covers only a finite region in $\dR^2$; the
result is drawn qualitatively in Fig.\ 2c.
The boundary on the right-hand side can be made straight by a conformal
transformation in $x^+$, by means of which one can also transform all
rectangles into squares. The final building block for the Penrose diagram
is obtained by turning the patch 45 degrees counter-clockwise, as before,
and is depicted in Fig.\ 2d.

In this example all the lines $X^3 \to \pm \infty$ are
complete: The null extremals running there are complete since the coordinate
$x^0$ ranges over all of $\dR$ in Fig.\ 2a (cf.\ the discussion following
Eq.\ \re{null2b})); for the non-null extremals this follows from \re{11}),
since $h(x^0)=O((x^0)^2)$. We will draw complete boundary lines boldfaced
and incomplete ones as thin solid lines. Horizons are drawn as dashed lines
(degenerate horizons, i.e.\ at higher order zeros of $h$, as multiply dashed
lines); the other null-extremals,
which run through the Penrose diagrams under $\pm 45$ degrees, are omitted.
Finally, massive dots indicate points at an infinite distance.

\newpage % Sorry, but LaTeX is too silly to place a
         % full-page figure correctly !!

{
%\begin{figure}[t]
\begin{center}
\leavevmode
\epsfxsize 12cm \epsfbox{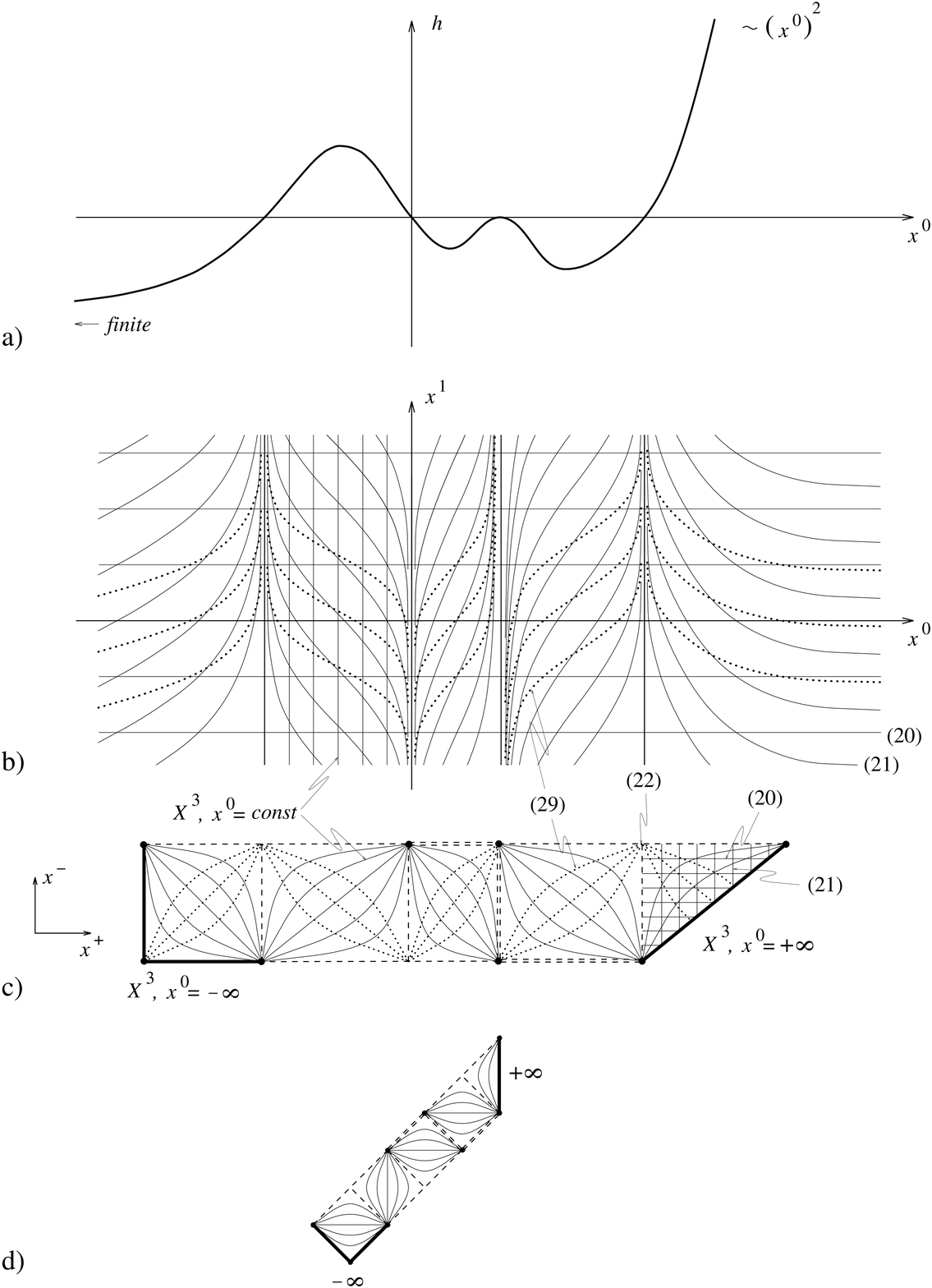} 
\end{center}
\renewcommand{\baselinestretch}{.9}
\small \normalsize
{\bf Figure 2:} {\small Construction of the fundamental building block for
a fictitious function $h(x^0)$. In (b) and part of (c) the null-extremals
\re{null1},\ref{null2a}) are drawn; in the final building block (d) they would
run under $\pm 45$ degrees, but have been omitted. In (c),(d) the horizons
\re{null2b}) have been drawn as dashed lines (degenerate ones as multiply
dashed lines). There and in the Penrose diagrams we have also included the
Killing trajectories (= lines of constant $X^3$), which in (b)
have been the straight lines $x^0= const$, as thin solid lines.
The function $X^3$, and simultaneously $x^0$, increases monotonically
throughout the block and the ``boundary'' values of $X^3$ are written to the
corresponding boundary segments. Finally, in (b),(c) we have drawn the
special extremals \re{13}), which are simultaneously the possible flip-axes
for the extension, as dotted lines.
% }
}
%\end{figure}
}

\newpage 

As pointed out before, each sector exhibits the flip symmetry. In the
example Fig.\ 2c,d we took this into account by drawing
the sectors symmetrically; especially those segments of the boundary depicted
as dashed lines are horizons, too, similar to the interior horizons.
The solution will thus be extendible at these boundary lines
and we must pay
attention that the conformal diffeomorphism transforming the $x^-$-coordinate
has the proper asymptotic behaviour to allow a smooth extension.
Due to the flip symmetry there is of course a distinguished choice for the
$x^-$-transition function, namely the one making the
sector symmetric (this amounts essentially to taking $x^-$ equal to 
some affine
parameter of an $x^+=const$ null-extremal). Unfortunately, different sectors
may yield contradictory functions, so in general one cannot simultaneously
achieve symmetry for all the sectors of the building block.

\begin{figure}%[h,t]
\begin{center}
\leavevmode
\epsfxsize 11 cm \epsfbox{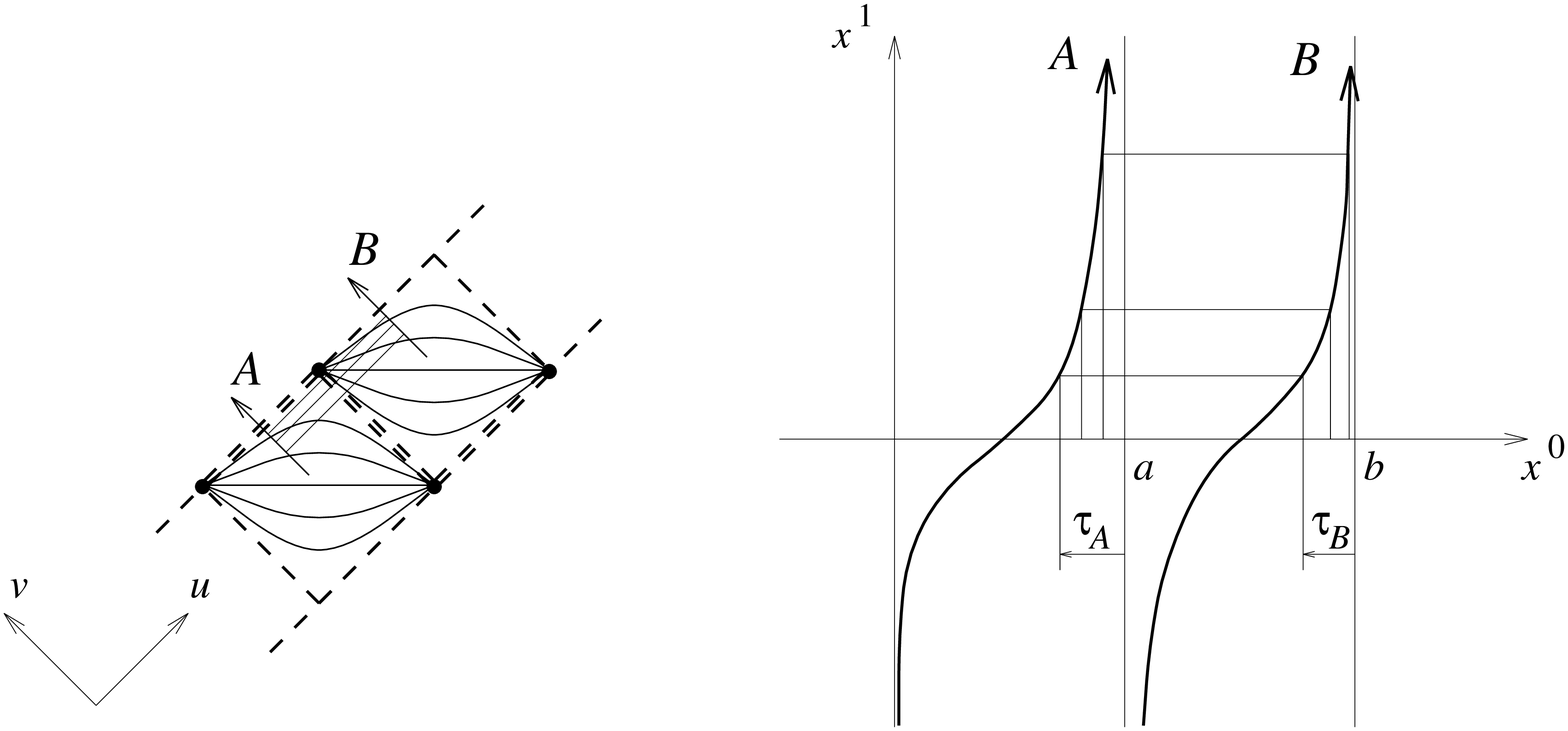}
\end{center}
\renewcommand{\baselinestretch}{.9}
\small \normalsize
{\bf Figure 3:} {\small Illustration of relative blue-shift.}
\end{figure}
The situation becomes particularly drastic
if $h(x^0)$ has zeros of different order as, e.g., in the example chosen in
Fig.\ 3 (cf.\ also Fig.\ 2 and the examples {\bf R3,4, G7,10} in
Sec.\ \ref{Recipe}):
Let $A$ and $B$ be null-extremals leaving the building block across a
degenerate (double zero of $h$) resp.\ non-degenerate horizon as shown in
Fig.\ 3. In the chart \re{011h}) they are described by Eq.\ \re{null2a})
and as their affine distance to the horizon we may choose $\t_A:=a-x^0$
and $\t_B:=b-x^0$. The conformity of the building block requires that a chosen
spacing along the null-extremal $A$ transfers to the extremal $B$ via the
family of perpendicular null-extremals \re{null1}), i.e.\ the lines
$x^1$ or $v=const$,
(cf.\ Fig.\ 3). Now, near the horizon the extremal $A$ may be described
asymptotically by $x^1\sim1/\t_A$ and $B$ by $x^1\sim-\ln\t_B$, hence
$\t_B\sim e^{-1/\t_A}$. When approaching the horizons $\t_A$ and $\t_B$
vanish simultaneously, but also ${d\t_B\0 d\t_A}\rightarrow 0$.
This already impedes a smooth diagram, which is seen as follows:
Choose a conformal gauge \re{confo}) and let $A$ and $B$ be represented
by $u=const=u_A$ or $u_B$, resp. The affine parameter along these
null-extremals then satisfies $d\t_A/dv \propto \exp(\r(u_A,v))$ and
likewise for $B$, as is easily shown. But from the above we know that
at the horizons
\be {d\t_B\0 d\t_A} = {d\t_B/dv\0 d\t_A/dv} \propto
  {\exp(\r(u_B,v))\0\exp(\r(u_A,v))} \rightarrow 0 \, .\nonumber \ee
Thus the conformal factor $\exp(\r(u,v))$ will either diverge along $A$ when
approaching the horizon or, if rendered finite along $A$, vanish along $B$.
It is thus {\em impossible\/} to draw a smooth Penrose diagram 
in this case! 
A physical consequence of this is that
an observer $B$ approaching the horizon, watching $A$ approaching the
degenerate horizon (both observers may also be timelike), will notice
that $A$ becomes infinitely blue-shifted.% FOOTNOTE
\footnote{More precisely, if $A$ is sending out 
  signals towards $B$ at a constant rate with respect to its affine
  parameter (proper time), $B$ will receive them with an ever increasing and
  finally diverging frequency (with respect to its own affine parameter).} 
Similar problems occur for any combination of different horizons; only
if all horizons are of equal degree the diagram will be smooth.

Nevertheless, it is still possible and instructive to use
such ``non-smooth Penrose diagrams'' for book-keeping when constructing the
extension and for studying the causal structure. In the illustrations
Fig.\ 11,\Figthirteen\ we have marked those non-smooth diagrams with the
sign \attn. At any rate, the extended solutions themselves will be smooth. 

But even for only non-degenerate horizons there may occur problems, if two 
triangular sectors meet. While for {\em one\/} sector the boundary can always
be made straight by a conformal diffeomorphism, this may be impossible for
{\em two\/} adjacent sectors simultaneously: 
Any conformal diffeomorphism (which is a reparametrization of each of the two
lightcone-coordinates $x^+,x^-$) alters the angles between boundary and
horizon on both sides of the horizon in the same sense; more precisely, the
ratio of the tangents of the respective angles is invariant.
\begin{figure}%[h,t]
\begin{center}
\leavevmode
\epsfxsize 14cm \epsfbox{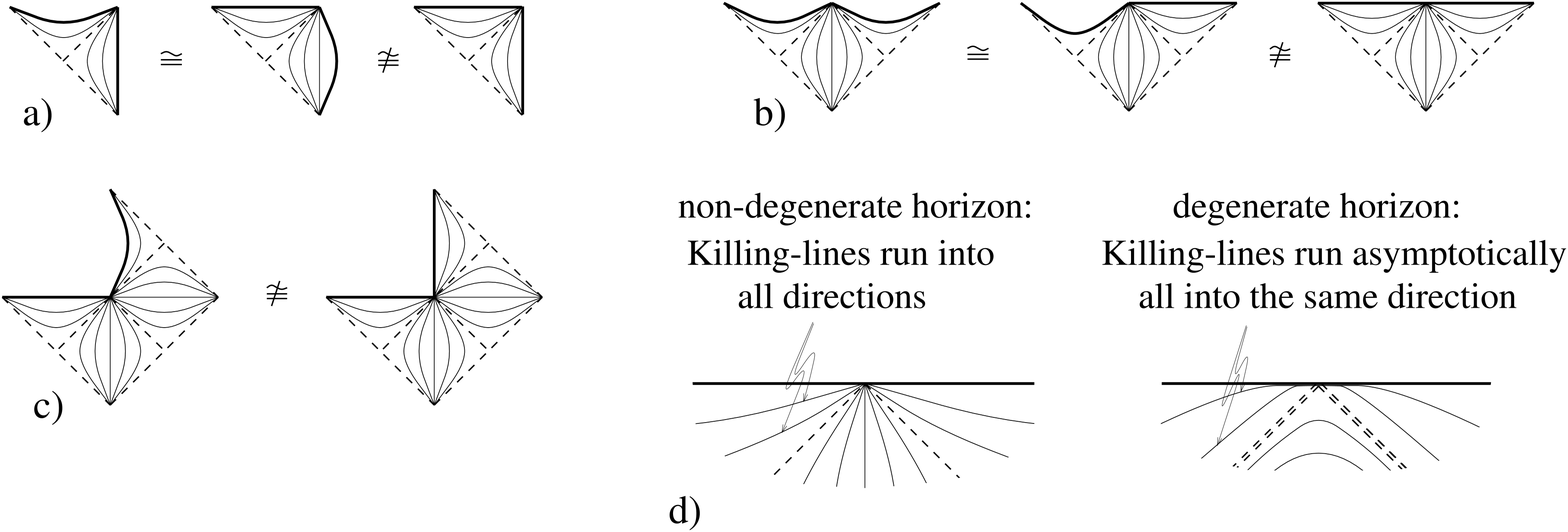}
\end{center}
\renewcommand{\baselinestretch}{.9}
\small \normalsize
{\bf Figure 4:} {\small Distorted Penrose-diagrams.}
\end{figure}
For instance, in the situation of Fig.\ 4a, a conformal diffeomorphism which
makes the upper boundary straight must necessarily bulge out the right
boundary (at least near the corner point), so the angles of the boundaries
against the horizon can never be made equal. This argument is also valid for
the (already ``extended'') diagram Fig.\ 4c, and for similar reasons the angle
in Fig.\ 4b cannot be smoothed. At the end of Sec.\ \ref{Saddle}
(cf.\ Fig.\ 7) we will show by an example that the distorted
diagram is in fact the generic case, if two ``triangular'' sectors meet. 

Finally, for degenerate horizons of equal degree these distortions do not
occur; one can always obtain right angles or straight lines. But also the
Killing lines behave differently then: As can be shown easily using
\re{011h},\ref{null1},\ref{null2a}) they no longer run from the corner
point of the sector in all directions; instead they leave the corner point in
only one direction asymptotically (tangential to the boundary, if there is
one; cf.\ Fig.\ 4d). This feature applies also to (4d) extremal
Reissner-Nordstr\"om with its degenerate horizons, which is in this
respect drawn incorrectly in most textbooks (e.g. \cite{Hawk}).

\section{Maximal Extension and Saddle-Point Charts}

\plabel{Saddle}

In the previous section we have compressed the coordinate patches \re{011h})
into finite building blocks. Now we will investigate 
whether these blocks have to be extended and how to do so.

Let us again examine the cases without zeros of $h$ first. 
Their building blocks are drawn in Fig.\ 1 and they are already inextendible,
which is seen as follows:
If an extension were possible at all, then it must be possible (also)
along the null-infinities (i.e., the 45-degree boundaries of Fig.\ 1a,b)
or along the timelike (for $h>0$) singularities% FOOTNOTE
\footnote{We will use this term even if the ``singularity'' turns out to be at
  an infinite distance.}
(the vertical resp.\ curved
boundaries in Fig.\ 1b,c). The extension cannot be possible over a corner
point {\em alone\/}, since this point then cannot be an interior point
of a new local chart.
%
%(At least we assume that such an ``extension'' is not
%desirable; perhaps it is actually impossible, ... don't know).
%

Now, if $x^0$ ranges over all of $\dR$, then these boundary
lines lie at an infinite distance since $x^0$ is an affine
parameter for the null-extremals \re{null1}, \ref{null2a}). 
Thus they cannot be regular interior points
and the solution is inextendible. Note, however, that in general null-
and non-null extremals may have different completeness properties!
The completeness of the null-extremals depends on the domain of $x^0$, 
whereas the completeness of the other extremals hinges largely on the
(asymptotic) behaviour of the function $h$ (cf.\ Eq.\ \re{11})).
E.g., the boundaries for the $R^2$-model \re{R2}) are
``null-complete'' (as $x^0 \in \dR$) but ``non-null incomplete'' 
(as $\lim_{x^0 \to \pm \infty} s =$ finite, cf.\ \re{11})). Usually a
boundary point is
called complete only if {\em all\/} extremals running into it are complete.
On the other hand, {\em one\/} complete extremal is sufficient to make the
metric inextendible through that boundary point.

If $x^0$ does not range over all of $\dR$, then the null-extremals are
incomplete and perhaps also the other extremals. However, even then the
solution is inextendible, since on these boundaries some physical field
($X^3$ or some dilaton field) diverges. Of course, if these fields are not
taken seriously (or if one drops the restriction $D'(\Phi) \neq 0$), then
one might try to extend the function $h(x^0)$ smoothly or analytically
beyond its original domain and repeat the analysis with this new $h$ to
obtain an extension.

In the case where $h$ has zeros, the discussion above is still valid for
the first and the last sector, i.e., an extension is not possible beyond
those boundary lines. However, at the other, interior, boundary lines the
solutions are extendible and this can be done as follows:
Already in Sec.\ \ref{General} we have introduced the flip transformation
\re{glue}). In the Penrose diagrams this flip shows up as a reflexion
of the sector in question.
The reflexion axis, running diagonally through
the sector, has to be {\em transversal\/} to the Killing lines
(i.e.\ horizontal for stationary sectors and vertical for homogeneous ones;
cf.\ e.g.\  the dotted lines in Fig.\ 2c). 
%Since this ``orientation'' of the sectors changes only at horizons of odd
%degree, the degrees of the zeros of $h$ will have a decisive impact on the
%extension of the solutions.
The flip transformation breaks down at the horizons bounding the
sector, because it maps the interior horizons onto exterior ones. 
We can, however, flip (i.e.\ reflect) the whole building block and thus obtain
an extension beyond the previously ``exterior'' horizons.
The gluing thus amounts to taking the mirror image of the block and patching
the corresponding sectors together.
This procedure has to be performed at each sector of the first building block
and after that also at each sector of the new building blocks and so on   
(cf.\ Fig.\ 5 for the first steps).
\begin{figure}[t]%[h,t]
\begin{center}
\leavevmode
\epsfxsize 14cm \epsfbox{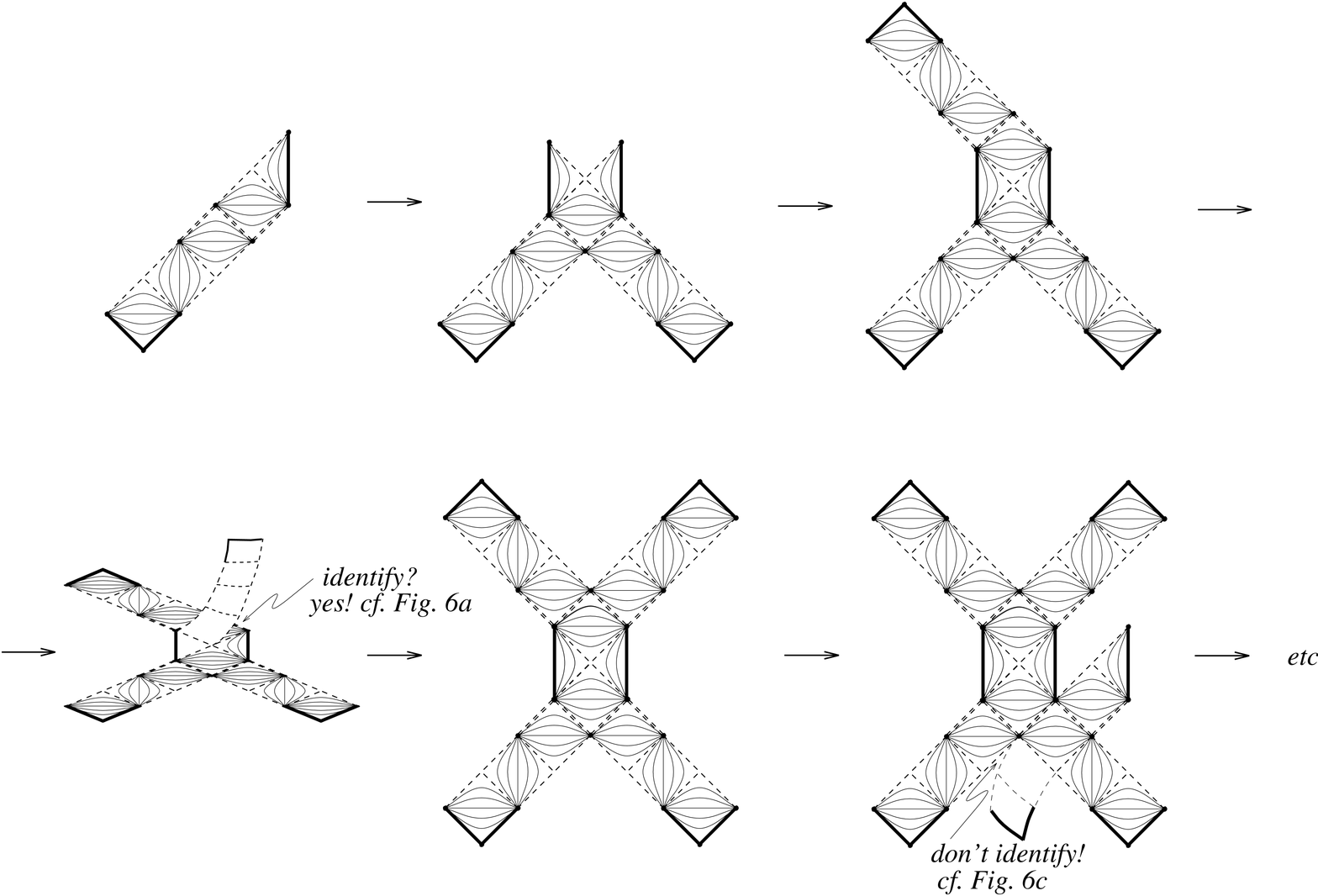}
\end{center}
\renewcommand{\baselinestretch}{.9}
\small \normalsize
{\bf Figure 5:} {\small The gluing procedure.}
\end{figure}

This gluing process is essentially unique; the only free parameter is the
choice of the symmetry axis of the flip transformation (i.e.\ the choice of
the constant in \re{fun})), but it only results in a coordinate change
(shifting the $x^1$-origin of the charts \re{011h})). As long as only the
universal covering is pursued this does not affect the solution. However,
when further identifications of sectors are made, such that the resulting
solution is not simply connected, then we can have an effect of this
parameter (this will be discussed in Part III \cite{partX}).

\medskip

We have yet to investigate the case (see Figs.\ 5,6)
that after surrounding the  point at the vertex of four blocks
the overlapping sectors of the first and the fourth block
match. Shall they be identified?

This depends: There is exactly one case where they must be identified
(Fig.\ 6a) and there are a couple of cases where this could (but shall not)
be done. A necessary requirement is of course that the sectors are equal
(i.e.\ isometric, scalar curvature and fields like $X^3$ coincide).
This is the case in (a),(d)
of Fig.\ 6, but generically not in (b),(c), because there different sectors
overlap (e.g.\ in (b) the sectors 1 and 5, if numbered consecutively in the
building block); only under rare circumstances (e.g.\ periodic functions $h$
and $X^3$) they could happen to fit onto one another. 

\begin{figure}[t]%[h,t]
\begin{center}
\leavevmode
\epsfxsize 11cm \epsfbox{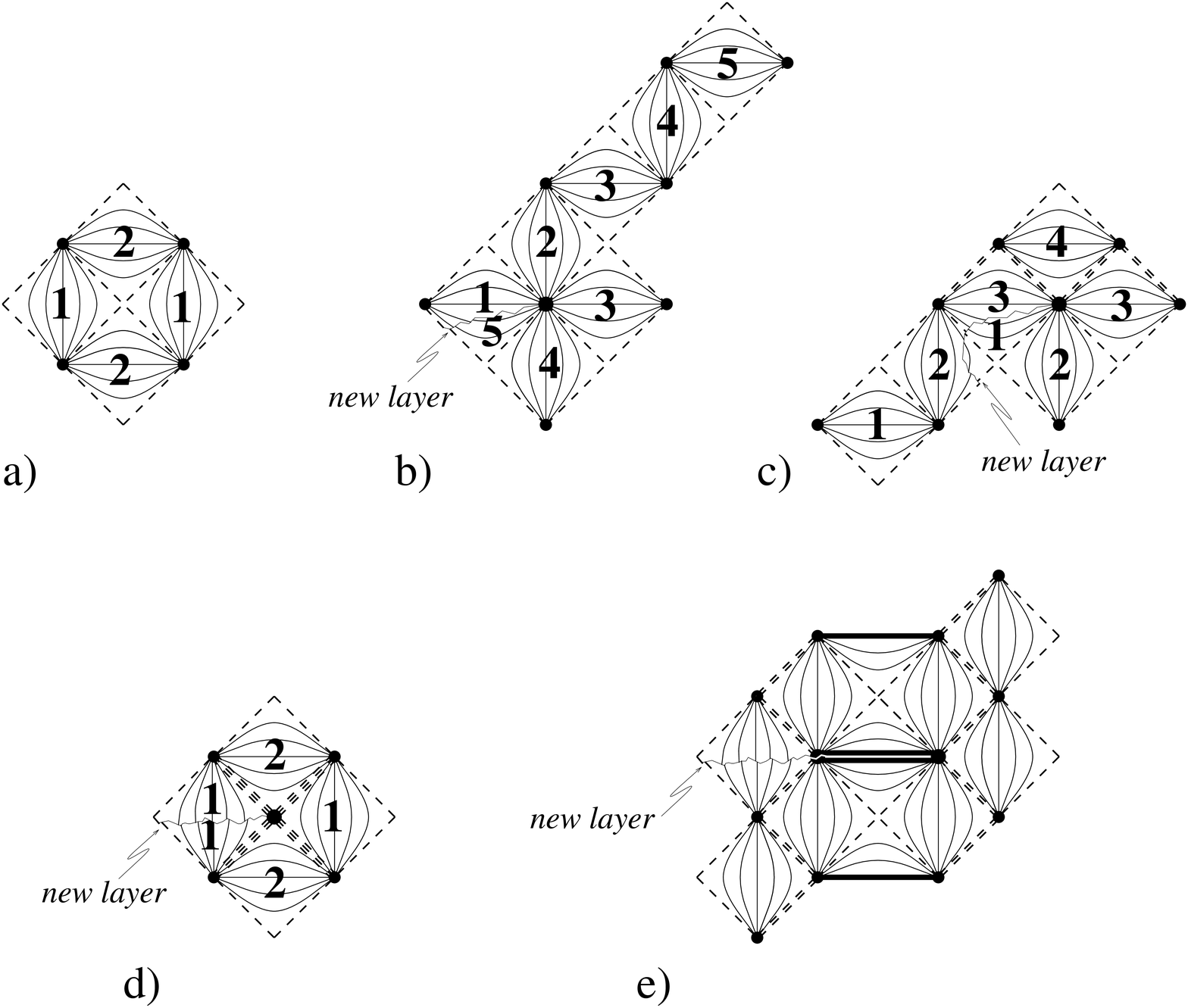}
\end{center}
\renewcommand{\baselinestretch}{.9}
\small \normalsize
{\bf Figure 6:} {\small Gluing around a vertex point. Only in (a) the four
sectors make up a single sheet around the vertex point. In all other cases
(b--d) the vertex point is not an interior point and the solutions must be
extended into further new layers covering the original sectors. The same is
true for the general situation of e.g.\ (e), showing part of the extended
diagram {\bf G7} (cf.\ Fig.\ \Figthirteen).}
\end{figure}

Now to the question whether this shall be done:
If the solution {\em can\/} be extended smoothly into the vertex point, then
of course this must be done, and in this case the overlapping sectors have
to be identified to make a single sheet around this vertex point (as shown
below this will be the case precisely in the situation Fig.\ 6a).
If, on the other hand, the vertex point turns out to lie at an infinite
affine distance, then it cannot possibly be an interior point. If under
these conditions the sectors were identified, then one would have a
non-contractible loop around this vertex point, which cannot be accepted
since we intend to construct the (simply connected) universal covering.
Thus we must not identify the overlapping sectors but start a ``new
layer" and continue the gluing, giving rise to the winding-staircase like
structure outlined in Fig.\ 6b--d (cf.\ also Fig.\ 11/{\bf R2} or
Fig.\ \Figthirteen/{\bf G4}). 
This applies also to the situation (Fig.\ 6e) that a whole slit is
surrounded before overlapping, or even more generally to any occurrence of
overlapping sectors different from the arrangement Fig.\ 6a.
On the other hand, if multiply connected solutions are allowed, then an
identification of overlapping sectors is possible, certainly.
In this case the gluing process is ambiguous, however, and introduces
a new geometrically meaningful parameter.
Such solutions will be analyzed in full detail in Part III.

Our considerations on the completeness of the horizons (cf.\ the paragraph
below \re{null2b})) show already that Fig.\ 6a is the only candidate for a
regular vertex point: Degenerate horizons are complete in both directions,
whereas non-deg.\ horizons are complete on one side only, which is easily
recognized as that side of the building block where the Killing lines converge
(in Fig.\ 6 all such complete points have been marked by
massive dots). But even the general extremals running towards those vertex
points show the same behaviour: In sectors where the Killing lines run into
the vertex point, these extremals are either oscillating ones
(between two fixed $X^3$- resp.\ $x^0$-values, cf.\ \re{exsol1})) or of the
kind \re{exsol2}), both of which are complete. In the other sectors, where
they have to run transversally to the Killing lines, these extremals are
precisely those of \re{exsol1}) with $c=0$, i.e.,
\be {dx^1\over dx^0}=-{1\over h},
  \el 13 
and coincide, furthermore, with the possible symmetry axes for the flip
transformations (in Fig.\ 2b,c we have drawn some of them as dotted lines).
Their length follows from \re{11}) where $a$ is the zero of $h(x^0)$ in
question:
\be s=\int^a {dx^0\over\sqrt{|h|}} \sim \int^a
  {dx^0\over{(x^0-a)^{n\over2}}} \to \left\{
                         \begin{array}{r@{\quad}l}
                                <\infty & n=1 \\
                                 \infty & n\ge 2
                         \end{array} \right. \quad. \el length
It is finite only at simple zeros. Thus, whenever a degenerate horizon runs
into the vertex point or whenever Killing lines
focus there, then this vertex point is at an infinite distance
and has to be taken out of consideration. In these cases, to obtain the
universal covering we must not identify the overlapping sectors but continue
the gluing in a new layer.

In the situation of Fig.\ 6a (only non-deg.\ horizons and the Killing lines
avoiding the vertex point), however, all extremals are incomplete, and it is
to be expected that the vertex point is an interior point of the
manifold. This is indeed true:
Eq.\ \re{saddle}) represents a smooth nondegenerate
metric for the neighbourhood of such points,
which reveals the vertex point as regular interior point
(a saddle point of $X^3$, with $X^a=0$).
The four adjacent sectors then constitute one single sheet (cf.\ also
diagrams 3--5 of Fig.\ 5).
%\footnote{In this case the constant of the last (fourth) gluing
%diffeomorphism \re{glue},\ref{fun})) is already uniquely determined by the
%requirement that the solution be smooth at the vertex point.}

Keeping in mind the Schwarzschild-like form of \re{gSS}), there is also an
alternative way of obtaining such a saddle-point chart, imitating the 
Kruskal-Szekeres procedure. It was initially
proposed by M. Walker \pcite{Walker} for a special rational form of $h$.
We want to show here that it works for {\em any\/} sufficiently smooth $h$
(say $C^n$) with simple zero at $x^0=a$.
The transformation reads (with $f$ as in \re{fun}))
\be u=\hbox{sgn}(h)e^{h'(a)\left[f(x^0)+\frac{x^1}{\scriptstyle2}\right]}
  =\hbox{sgn}(h)e^{\int\limits^{\,\,\,\,x^0}\!\!\frac{\scriptstyle 
  h'(a)}{\scriptstyle h(x)}dx
  +\frac{\scriptstyle h'(a)}{\scriptstyle2}x^1 }
  \el Krtrfu
\be v=e^{-\frac{\scriptstyle h'(a)}{\scriptstyle2}x^1}  \, \, .
  \el Krtrfv
It brings the metric \re{011h}) into the form
\be ds^2=-\frac{4\mytilde h(uv)}{h'(a)^2}\frac{dudv}{uv}\, ,
  \el Kruskal
which is evidently nonsingular and nondegenerate around $u,v=0$.  
Here $\mytilde h(uv)$ is determined  implicitly 
via $\mytilde h(uv)=h(x^0(u,v))$. 
The integration constant hidden in $u$ has to be chosen
such that $u$ is continuous at $x^0=a$. 

To show that the transformation \re{Krtrfu},\ref{Krtrfv}) is really a
(local) diffeomorphism
consider the following decomposition of the integrand in \re{Krtrfu})
(separating the singular term):
\be \frac{h'(a)}{h(x)}=\frac1{x-a}-\frac{\frac{h(x)}{x-a}-
  h'(a)}{h(x)}\,\, . \ee
The first term yields (after integration and exponentiation) a factor
$(x-a)$,% FOOTNOTE
\footnote{Note that the coefficient $h'(a)$ in \re{Krtrfu}) is absolutely
  necessary: Otherwise we would have a factor $k\ne1$ in the singular term,
  which then integrates to $(x-a)^k$, and the transformation would no longer
  be a diffeomorphism. This seems to have been ignored in the literature
  sometimes.}
and also the second term 
(being $C^{n-2}$, if $h\in C^n$) behaves
perfectly well. It is thus not at all necessary that $h$ is of the special
rational form given in \pcite{Walker,Brill}.

\medskip

We are now finally in the position of proving the statement on the 
distorted boundary of
the $R^2$-gravity solutions made at the end of Sec.\ \ref{Blocks}.
Let e.g.\ $\L=-{1\03},C={4\03}$, hence
(cf.\ Eq.\ (\ref{R2})) $h={2\03}\left[2-x^0-(x^0)^3\right]$.
Eqs.\ \re{Krtrfu},\ref{Krtrfv}) can be integrated easily to
\ba uv&=&\mbox{sgn}(h)\exp\int\limits^{\,\,\,\,x^0}{{h'(1)}\0{h(x)}}dx=
                                                               \nonumber \\
      &=&\mbox{sgn}(h)\sqrt[8]{(x^0)^2-2x^0+1\0(x^0)^2+x^0+2}
  \exp\left({{-3\04\sqrt{7}}\arctan{1+2x^0\0\sqrt{7}}}\right)\,\, .  \ea
The singularities lie at $x^0=\pm\infty$, i.e.\ at 
$uv=\exp\left(-{3\p\08\sqrt{7}}\right)\approx1.561$ and
$uv=-\exp\left(+{3\p\08\sqrt{7}}\right)\approx-0.641$,
respectively. 
\begin{figure}%[h,t]
\begin{center}
\leavevmode
\epsfxsize \textwidth \epsfbox{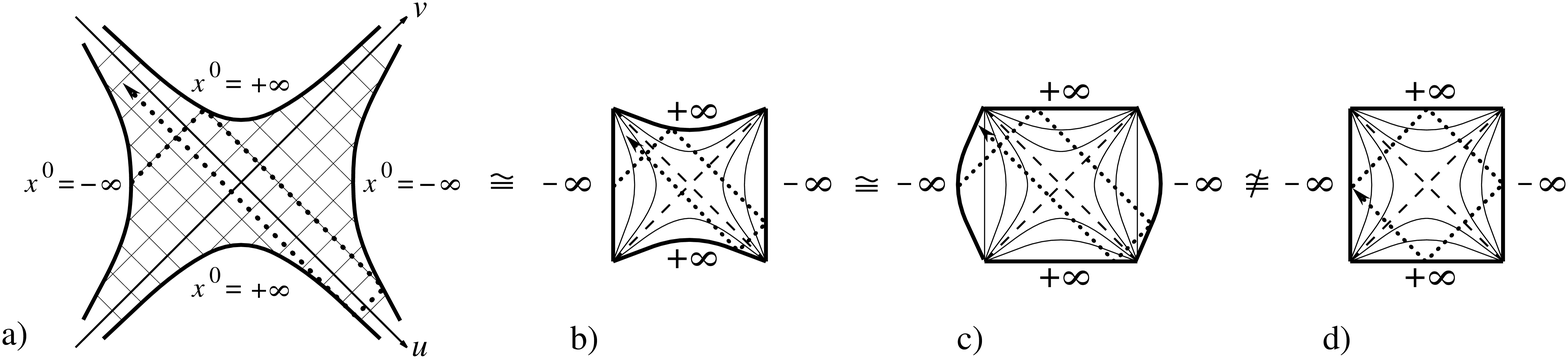}
\end{center}
\renewcommand{\baselinestretch}{.9}
\small \normalsize
{\bf Figure 7:} {\small Distorted Penrose-diagrams for the $R^2$-model.}
\end{figure}
The familiar conformal diffeomorphism $u \rightarrow \tan \hat u,
v \rightarrow \tan \hat v$ shows immediately that
the correct shape of the Penrose diagram must be like in Fig.\ 7b,c
(it is of course possible to straighten the one boundary line at the cost of
the other by means of conformal diffeomorphisms).
Only if the two hyperbolae $x^0=\pm\infty$ happen to lie at equal values of
$|uv|$ then we can obtain a square.
Thus for non-symmetric functions $h$ (with one zero) squares are rather the
exception to the rule. This remark applies also to the square shaped Penrose 
diagrams obtained, e.g., in \cite{Brill,Lemos} for other special models
covered by the present treatment. 
Finally we want to mention some physical difference between a world with a 
square-shaped Penrose diagram and another of the generic form Fig.\ 7b,c:
Only in the former case 
a series of null-lines ``bouncing off'' the boundaries 
closes to a rectangle (cf.\ dotted lines in Fig.\ 7).

\section{Recipe and Examples}

\plabel{Recipe}

Let us summarize briefly the principle of how to construct the maximally
extended Penrose diagram corresponding
to any function $h$ in \rp 011h :\\
$\bullet$ The number and kind of zeros of the function $h$ determines the
number of sectors and their ``orientation'' in a fundamental building block
($h>0$ corresponds to stationary sectors, $h<0$ to homogeneous ones). 
More explicitly: For $n$ zeros of $h$ there are
$n+1$ sectors in the building block. If a zero of $h$ is of odd
order, then the corresponding Killing horizon separates a stationary from a
a homogeneous sector; otherwise the two neighbouring sectors are clearly of the
same type (as $h$ does not change its sign).
\\
$\bullet$ The end sectors of the building block are either a square or a
triangle, depending on the asymptotic behaviour of $h$. It is a triangle,
if $f(x^0)=\int^{x^0} du/h(u)$ remains finite at the boundary (e.g., for
$h(x^0)\sim (x^0)^{k>1}$ as $x^0\to\pm\infty$), and it is a square, if
$f(x^0)$ diverges (e.g., for $h(x^0)\sim (x^0)^{k\le1}$
as $x^0\to\pm\infty$). In the case of a triangle the hypotenuse runs
parallel with the Killing lines, of course (i.e.\ vertical for stationary
sectors, horizontal for homogeneous ones).
\\
$\bullet$ Choose any sector and establish the symmetry axis, running
diagonally through the sector, transversal to the Killing lines
(i.e.\ horizontal for stationary sectors and vertical for homogeneous ones). 
Reflect the whole block at this symmetry axis and identify the corresponding
sectors.
\\
$\bullet$ Proceed in this way with all sectors until you come to an end, or
ad infinitum.
\\
$\bullet$ If after surrounding a vertex point sectors overlap, the Killing
horizons running into that point being non-degenerate (simple
zero of $h$), and if the Killing lines do not focus there (Fig.\ 6a),
then identify the overlapping sectors such as to make a single sheet around
this vertex point. In all other cases (e.g.\ higher degree zero of $h$,
Killing-lines running into that point, etc.) do not identify the
sectors, but continue gluing in a new layer to get the universal covering.
\\
$\bullet$ Any boundary  is null-complete, iff $x^0 \to \pm \infty$ there. 
A boundary point is complete with respect to all other extremals
\re{exsol1}), iff $\int^{x^0} du/\sqrt{|c-h(u)|}$ diverges there,
(cf.\ Eq.\ (\ref{11})).
Complete boundaries have to satisfy both conditions (except that only
extremals of one kind run there) and are depicted  boldfaced in the present
paper. 

As pointed out repeatedly before, the Penrose diagrams obtained in this 
way are to be understood as {\em schematic\/}  
ones only. However, if the zeros of the respective function $h$ are 
all of the same order,
then these diagrams are also 
smooth; by this we mean that there exists a (smooth) diffeomorphism 
from the universal covering solution to the respective diagram.
Still, if the zeros are all simple (non-degenerate horizons) and two
non-null boundary lines meet in a point, then it will in general be
necessary to deform the boundary lines somewhat (cf.\ Figs.\ 4a--c, 7).
For a null and a non-null boundary line meeting% FOOTNOTE
\footnote{If, say, a spacelike boundary meets a null boundary at a corner
  point (e.g.\ in the diagrams {\bf G1-4, 8,9,11} of Fig.\ \Figthirteen), then
  it is obviously no problem to change the angle of the spacelike boundary
  at will, since the null boundary always remains null under conformal
  transformations.}
or for higher order zeros these complications do not occur.

If, on the other hand, $h$ has zeros of different degree, 
then there is {\em no} smooth diffeomorphism from the universal covering 
solution to the Penrose diagram; the latter has to be regarded as purely 
schematic then. An explanation from the physical point of view has been
provided by means of a relative red/blue-shift between observers approaching 
Killing horizons of a different type (cf.\ Fig.\ 3). 
Certainly the universal covering solutions themselves are still 
smooth everywhere  in these cases, too: The gluing diffeomorphism \re{glue}) 
may be used to patch together charts \re{011h}) (this poses no problem since
only {\em one} kind of Killing horizon at a time is involved) and an 
atlas is obtained when completing  these charts by those of type \re{saddle})
or \re{Kruskal}). Obviously, also in these cases it is instructive to keep
track of the patching by a (schematic Penrose) diagram; only  it may not 
be taken as a true smooth image of the respective spacetime then. 

\vspace{.3cm}

We now come to the announced examples, starting with (anti-)deSitter gravity
(\ref{JT}). Since all values $\L \neq 0$ yield (basically) equivalent
Penrose diagrams, we will set $\L:=-2$ in the following. This describes
deSitter gravity (note that the sign of the curvature $R\equiv\L$ depends on
the signature convention, which in the present paper is different from e.g.\
\cite{Hawk}). For positive values of $\L$ (anti-deSitter gravity) the
Penrose diagrams obtained below have to be turned by an angle of 90 degrees,
while $\L=0$ yields the diamond like Minkowski space diagrams, of course.
With the above choice $\L:=-2$ we get a one-parameter family of functions $h$
[cf.\ Eq.\ (\ref{JT})], $h^{JT}(x^0)=C - (x^0)^2$, parametrized by the Casimir
constant $C = X_aX^a + (X^3)^2$ (cf.\ Fig.\ 8), which leads to the following
three qualitatively distinct cases:

\begin{figure}%[h,t]
\begin{center}
\leavevmode
\epsfxsize 6cm \epsfbox{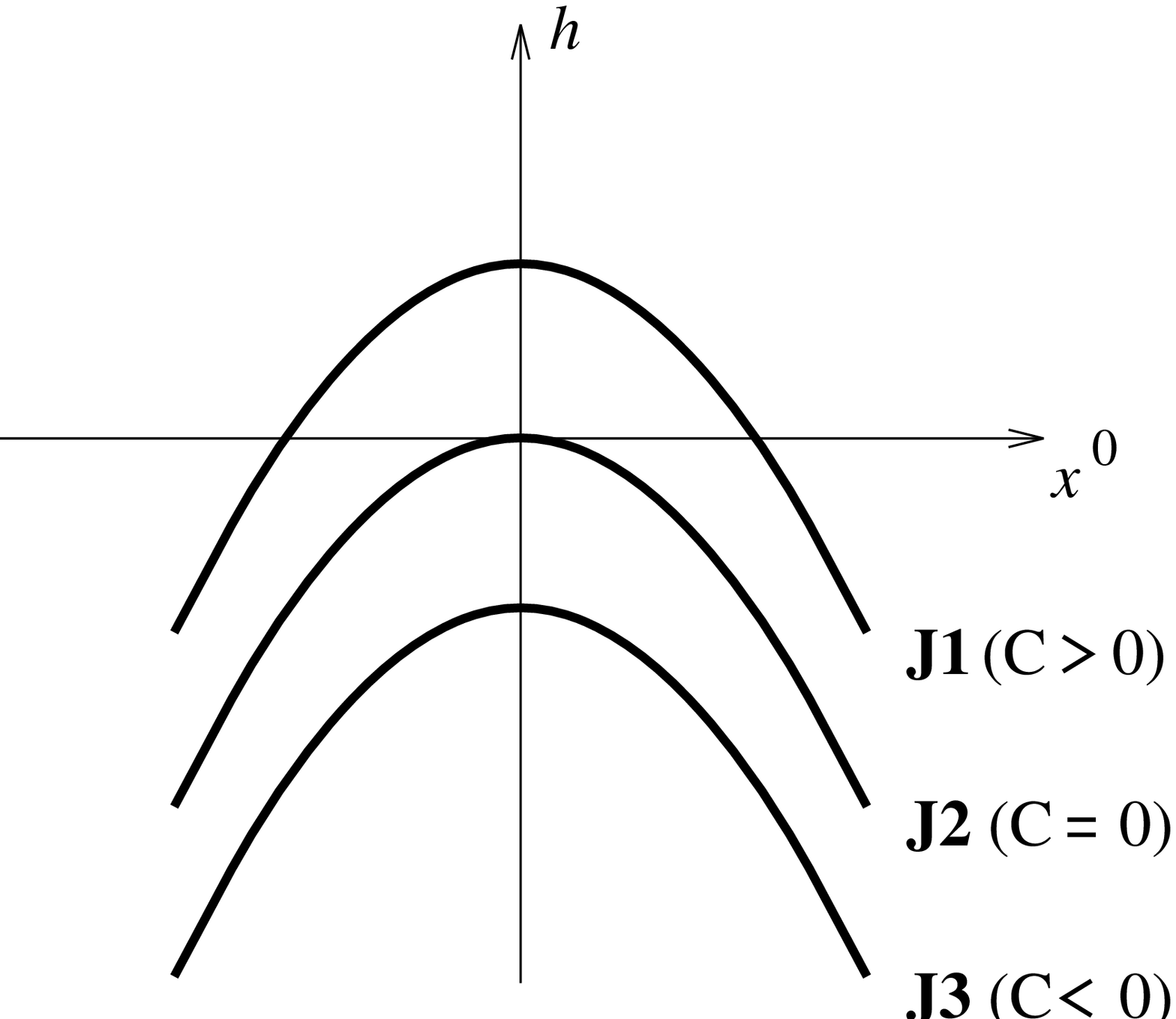}
\end{center}
\renewcommand{\baselinestretch}{.9}
\small \normalsize
{\bf Figure 8:} {\small Functions $h(x^0)$ for the JT-model (deSitter
gravity).}
\end{figure}

%\newpage % for DIN A4-Format

{\bf J1:} For any $C>0$ $h(x^0)$  has two simple zeros, leading to
one square within the fundamental building block.
Asymptotically we have $h\sim-{(x^0)}^2$, so that
adjacent to the square there will be a triangle at each side, the boundary
of which, $X^3 = \pm\infty$ resp., is spacelike and
complete. Gluing leads to the ribbon-like diagram shown in Fig.\ 9.
(For the other sign of the cosmological constant, e.g.\ $\L=+2$, we get the
same diagram for $C<0$, rotated, however,  by 90 degrees, as the boundary
is timelike then).

{\bf J2:} For $C=0$ we get no square, but only two triangles. The
corresponding Penrose diagram is again plotted in Fig.\ 9.

{\bf J3:} For $C<0$ the function $h$ has no zeros (Fig.\ 8). We therefore may
apply directly the diffeomorphism (\ref{conf})  with the function
(\ref{fun}), which in the present case can be written in terms of
elementary functions:
\be f(x^0)=-{2 \0 \sqrt{-C}} \arctan \left( {x^0 \0 \sqrt{-C}} \right) \, .
  \el JTfun
The resulting region $(x^+,x^-)$ is again a ribbon (Fig.\ 9);
but this time without such a periodic internal structure as before,
since the Killing lines $X^3=x^0=const$ become the parallels $x^+-x^-=const$
in the present case.

\begin{figure}%[h,t]
\begin{center}
\leavevmode
\epsfxsize 13 cm \epsfbox{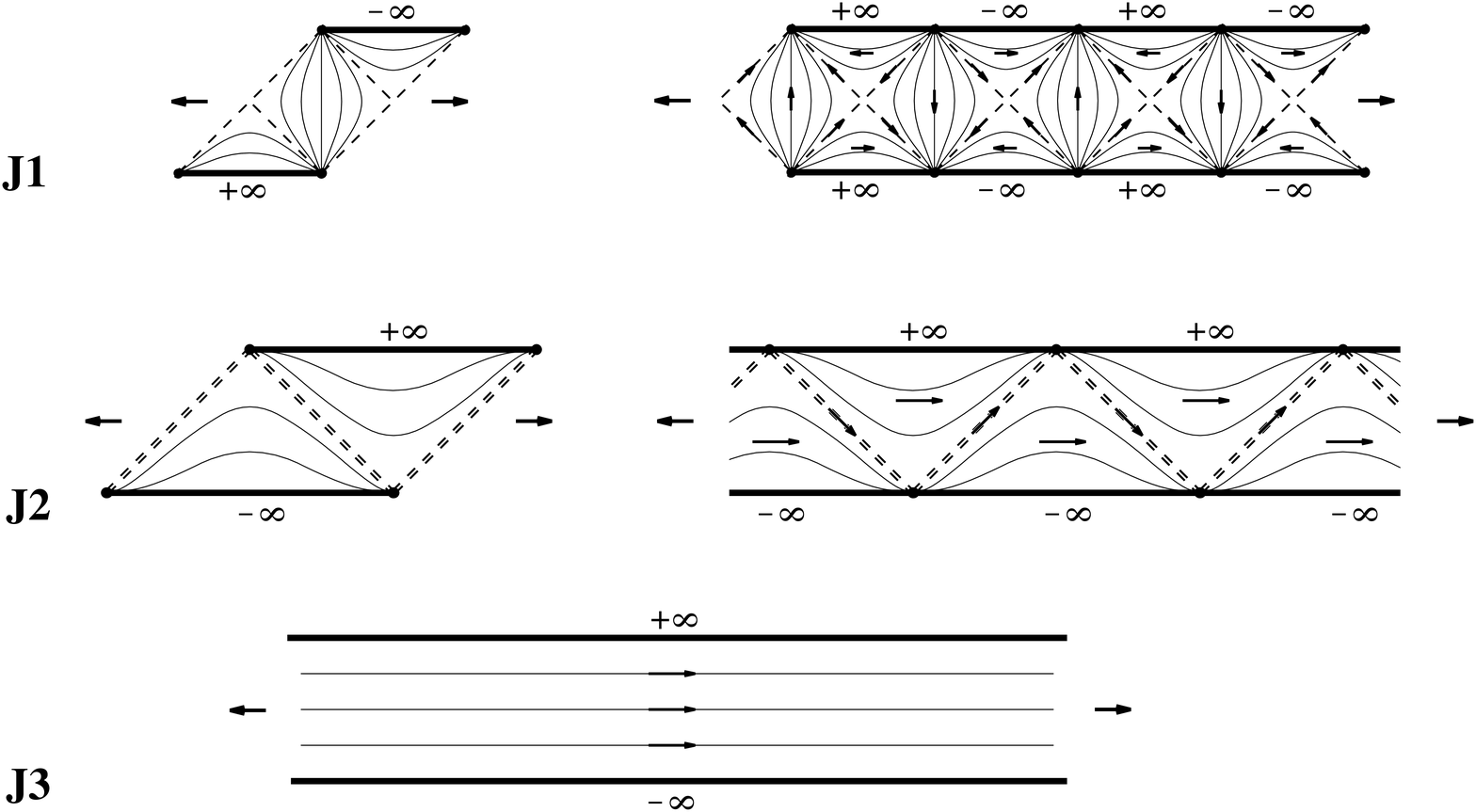}
\end{center}
\renewcommand{\baselinestretch}{.9}
\small \normalsize
{\bf Figure 9:} {\small Penrose diagrams for the JT-model; left the building
blocks, right the extended diagrams. Arrows inside the diagrams indicate
Killing fields for those solutions.}
\end{figure}

Clearly, as manifolds the solutions {\bf J1-3} are all the same, namely
the unique (simply connected, maximally extended) manifold with zero
torsion and constant curvature ($R\equiv\L$, cf.\ Eq.\ \re{JT})).
%
%Therefore the Penrose diagrams depicted in Fig.\ 9 are 
%in this case also smooth and not only schematic.
%
The difference between the spacetimes {\bf J1-3} 
arises only from the function $X^3$ defined on them. As mentioned already 
in Sec.\ \ref{Review}, preservation of $X^3$ reduces the originally three
independent Killing-fields to only one symmetry direction (indicated by
arrows in Fig.\ 9).

Note that while the Penrose diagrams for anti-deSitter space look the
same in any spacetime-dimension $D$ (vertical ribbon), this is not the
case for deSitter. There a ribbon-like structure appears for
$D=2$ only.
This may be understood as follows: $D$-dimensional deSitter space can be
obtained from restricting a $D+1$-dimensional Minkowski metric to the
one-sheet hyperboloid, the topology of which is $\dR\times S^{D-1}$.
For $D>2$ this hyperboloid is simply connected and thus it provides already
the universal covering space. Upon spherical reduction its Penrose
diagram is the familiar square \cite{Hawk,Thi}.% FOOTNOTE
\footnote{Precisely: After some transformations
  $g={const \0 \sin^2 t}[dt^2- d\O_{D-1}^2]$ (globally), where
  $d\O_{D-1}^2$ is the standard metric on the $D-1$-sphere.
  Extracting also the ``latitude'' $\c$ of the $S^{D-1}$,
  $d \O_{D-1}^2 = d\c^2+\cos^2\!\c\,d\O_{D-2}^2$, the space can be described
  as an $S^{D-2}$-bundle over the square $t \in (0,\pi)$,
  $\c \in [-{\p\0 2},{\p\0 2}]$.
  Note that the boundary lines $\c = \pm{\p\0 2}$ of this square are actually
  internal lines of the manifold (corresponding to the time-evolved North- and
  South-Pole of the $S^{D-1}$).}
For $D=2$, on the other hand, this hyperboloid is topologically a cylinder
$\dR\times S^1$ and the universal covering, obtained by unwrapping the
$S^1$ to $\dR$, is the horizontal ribbon of Fig.\ 9.
In comparison, to obtain a $D$-dimensional anti-deSitter metric,
one starts from the unit hyperboloid in $\dR^{2,D-1}$. This hypersurface has
topology $S^1 \times \dR^{D-1}$ and the (timelike) $S^1$ has to be unwrapped
for {\em any\/} dimension $D$, leading to the vertical ribbon diagram.
Concluding we also observe that for $D=2$ (and only there) deSitter and
anti-deSitter space are related by a simple signature change and the
corresponding Penrose diagrams are just rotated by 90 degrees against one
another.

Finally, it is of course possible to map the above infinite ribbons into
a finite region by a further conformal transformation. For anti-deSitter
space this leads for instance to the lens-like form of Fig.\ 1c and for
deSitter analogously to a horizontal lens. However, the periodicity of e.g.\
the solutions {\bf J1,2} is less obvious in this representation.
This remark is also valid for the other infinitely (periodically) extended
diagrams to be encountered below (cf.\ Figs.\ 11,13,14).
In some of those cases (e.g.\ {\bf R3-5, G7,11}) it is even possible to get
rid of the multilayered structure,
at the cost of introducing a fractal boundary (roughly speaking, the
overlapping sheets can be ``compressed'' into smaller non-overlapping
patches).

\medskip

\begin{figure}[t]%[h,t]
\begin{center}
\leavevmode
\epsfxsize 12cm \epsfbox{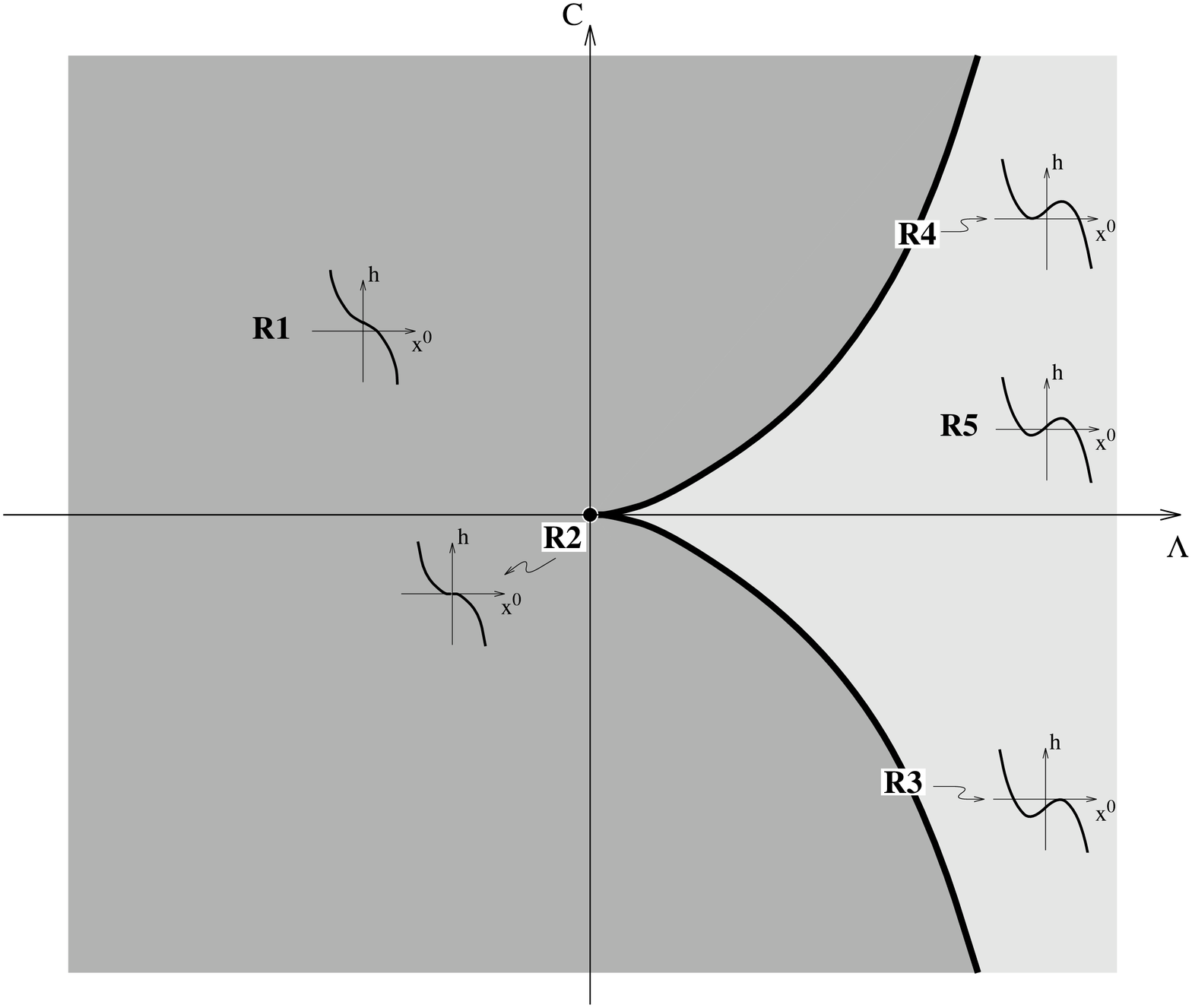}
\end{center}
\renewcommand{\baselinestretch}{.9}
\small \normalsize
{\bf Figure 10:} {\small $R^2$-gravity, overview.}
\end{figure}

Let us now turn to the second example: $R^2$-gravity \re{R2}). The
corresponding function $h(x^0)$ is in this case [cf.\ Eq.\ \re{R2})] 
$h^{R^2}=-\frac23(x^0)^3+2\L x^0+C$. Since $x^0$ ranges over all of $\dR$ and
at infinity $h\sim (x^0)^3$ we get incomplete (null-complete but
non-null-incomplete) triangular sectors at both ends of
the building block. The corner points are, however, complete, since only the
null extremals \re{null2b}) run into them.
A more detailed analysis yields five  qualitatively
different cases depending on the parameters $C$ and
$\Lambda$ (see Fig.\ 10):

\smallskip
\begin{tabular}{rl}
{\bf R1\ :}&one simple zero of $h$ at $x^0=B$ \\
{\bf R2\ :}&one triple zero at 0 \\
{\bf R3\ :}&one simple zero at $B_1$ and one double zero at $+\sqrt\Lambda$ \\
{\bf R4\ :}&one double zero at $-\sqrt\Lambda$ and one simple zero at $B_3$ \\
{\bf R5\ :}&three simple zeros at $B_1$, $B_2$, and $B_3$,
\end{tabular}

\smallskip
\noindent where
$B_1<-\sqrt\Lambda<B_2<+\sqrt\Lambda<B_3$ and $-\infty<B<+\infty$.
Obviously ${X^3}_{crit}=\pm \sqrt{\L}$ (by its definition as zero 
of the potential $W(0,X^3)$) and the curve  along {\bf R4,2,3} in
Fig.\ 10 corresponds to the critical values $C_{crit} \equiv
\pm (4/3)\Lambda^{(3/2)}$ of $C$.
It is completely straightforward to construct the  Penrose
diagrams according to the above rules. The result is depicted in
Fig.\ 11. Note that not only {\bf R2} but also the extended solutions
{\bf R3-5} will be multi-layered; for instance, the copies appended to
{\bf R4} at the upper left and the upper right horizon, resp., constitute
different layers of the universal covering (cf.\ Fig.\ 6e for a similar
situation).
For non-negative $\L$ there are, in addition to those diagrams, also
infinite ribbons for the  constant curvature solutions $d\o = \mp 2 
\sqrt{\L} \e$
(cf.\ Eq.\ \re{deSitter})); in the diagram Fig.\ 10 they are located at the
curve  {\bf R4,2,3}, i.e.\ at $C=C_{crit}(\L)$.

\begin{figure}[t]%[h,t]
\begin{center}
\leavevmode
\epsfxsize 11.5cm \epsfbox{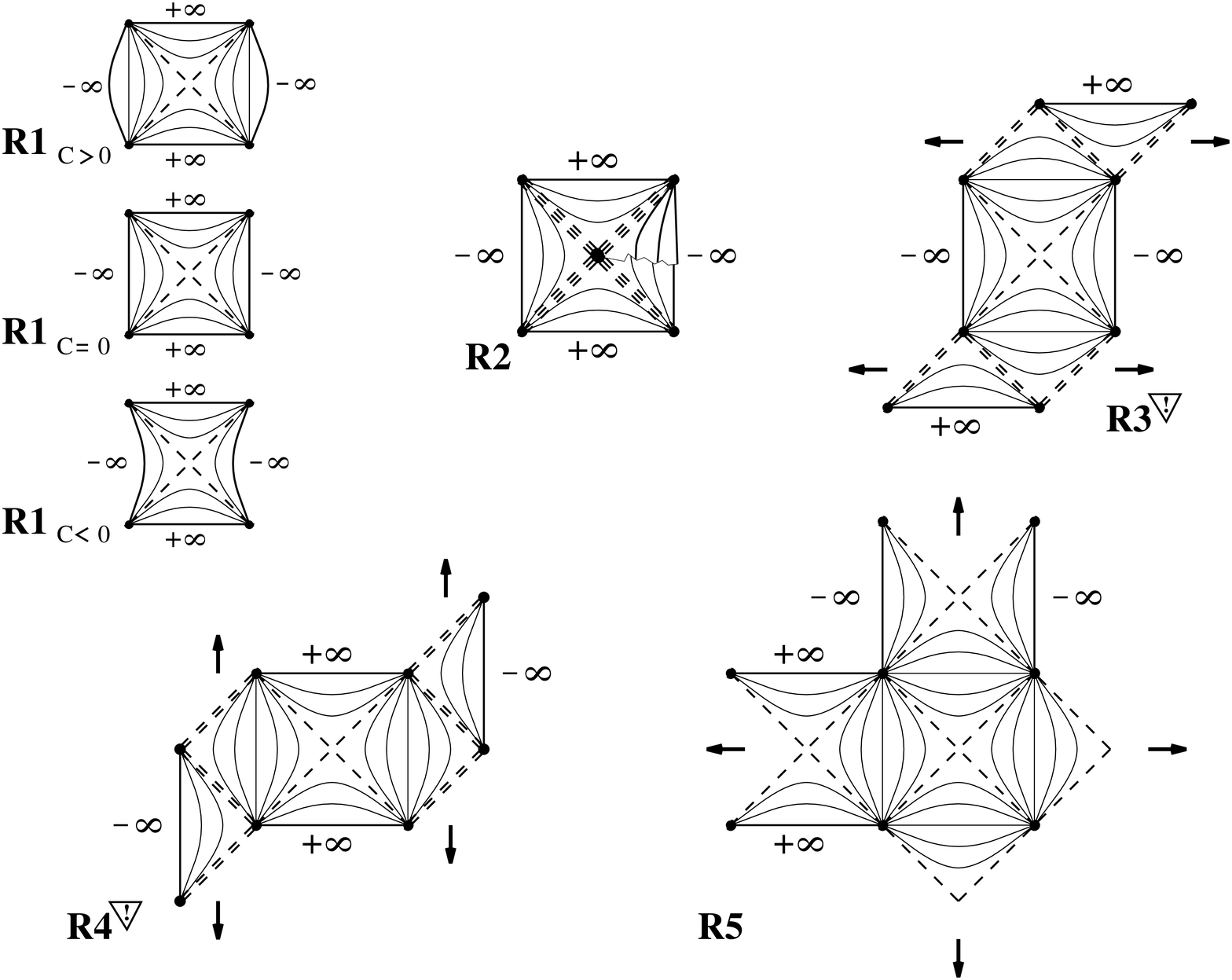}
\end{center}
\renewcommand{\baselinestretch}{.9}
\small \normalsize
{\bf Figure 11:} {\small Penrose diagrams for the $R^2$-model. Those marked
\attn \ \,are non-smooth; furthermore, the singularities of {\bf R5} might be
distorted in the same way as sketched in Fig.\ 4c. (For more details
cf.\ caption of Fig.\ \Figthirteen).}
\end{figure}

So, for a negative cosmological constant $\L$, 
which in contrast to $C$ is fixed by the choice 
of the action \re{R2}), there is 
only one schematic Penrose diagram (cf.\ Fig.\ 10). 
The choice of the one free parameter $C$, however,
influences the causal structure of {\bf R1} somewhat, giving rise to the
distorted boundaries (cf.\ also Fig.\ 7).
For positive values of $\L$ there are 
mainly two kinds of (again schematic) Penrose 
diagrams: {\bf R1}  and the a bit more complicated diagram 
{\bf R5}. Again, to have them smooth, some of the 
boundaries will be curved for generic values of $C$ (also for {\bf R5},
although this has not been indicated in Fig.\ 10!).
The separation of the solution space into solutions of the type {\bf R1} and 
{\bf R5} occurs at the values $\pm (4/3)\Lambda^{(3/2)}$ of $C$. At these 
values we have the spacetimes corresponding to the (inherently non-smooth) 
Penrose diagrams {\bf R3,4} {\em and\/} those of deSitter type with $R \equiv 
\pm 4 \sqrt{\L}$ (but now with no internal 
structure as the functions $X^i$ are constant all over $M$ here). 
So at $C=C_{crit}$ the 
Casimir constant does no longer classify the universal coverings uniquely. 
This is related to the fact that there is more than one symplectic leaf 
for a critical value of $C$: Besides a two-dimensional leaf, corresponding 
to {\bf R4} resp.\ {\bf R3}, there is the 
pointlike one $X^+=X^-=0$, $X^3=X^3_{crit}$, corresponding to a deSitter 
solution of positive resp.\  negative constant curvature. 

\medbreak The third example is the Katanaev-Volovich model \re{KV}). Its
function is
$h^{KV}(x^0) = {1 \0 \a}  \left\{ C x^0 - 2 (x^0)^2 [(\ln x^0-1)^2+1-\Lambda]
\right\} \,,$
where $x^0 \in \dR^+$.
The function $h^{KV}$ always has a  zero at $x^0 =0 \LRA \a X^3 = -
\infty$, which is at least of order one. Thus the function 
$f=\int^{x^0}du/h(u)$ (cf.\ Eq.\ \re{fun})) has infinite range as
$x^0\to 0$, 
which shows that the Penrose diagrams have
a square-shaped sector at this end, the null-infinities, however, being
incomplete. On the other boundary of the
coordinate patch, $\a X^3 = + \infty$, we find $\a h \sim
(x^0)^2 \ln^2 x^0$, corresponding to a complete triangular sector at this end.

\begin{figure}[t]%[h,t]
\begin{center}
\leavevmode
\epsfxsize 12cm \epsfbox{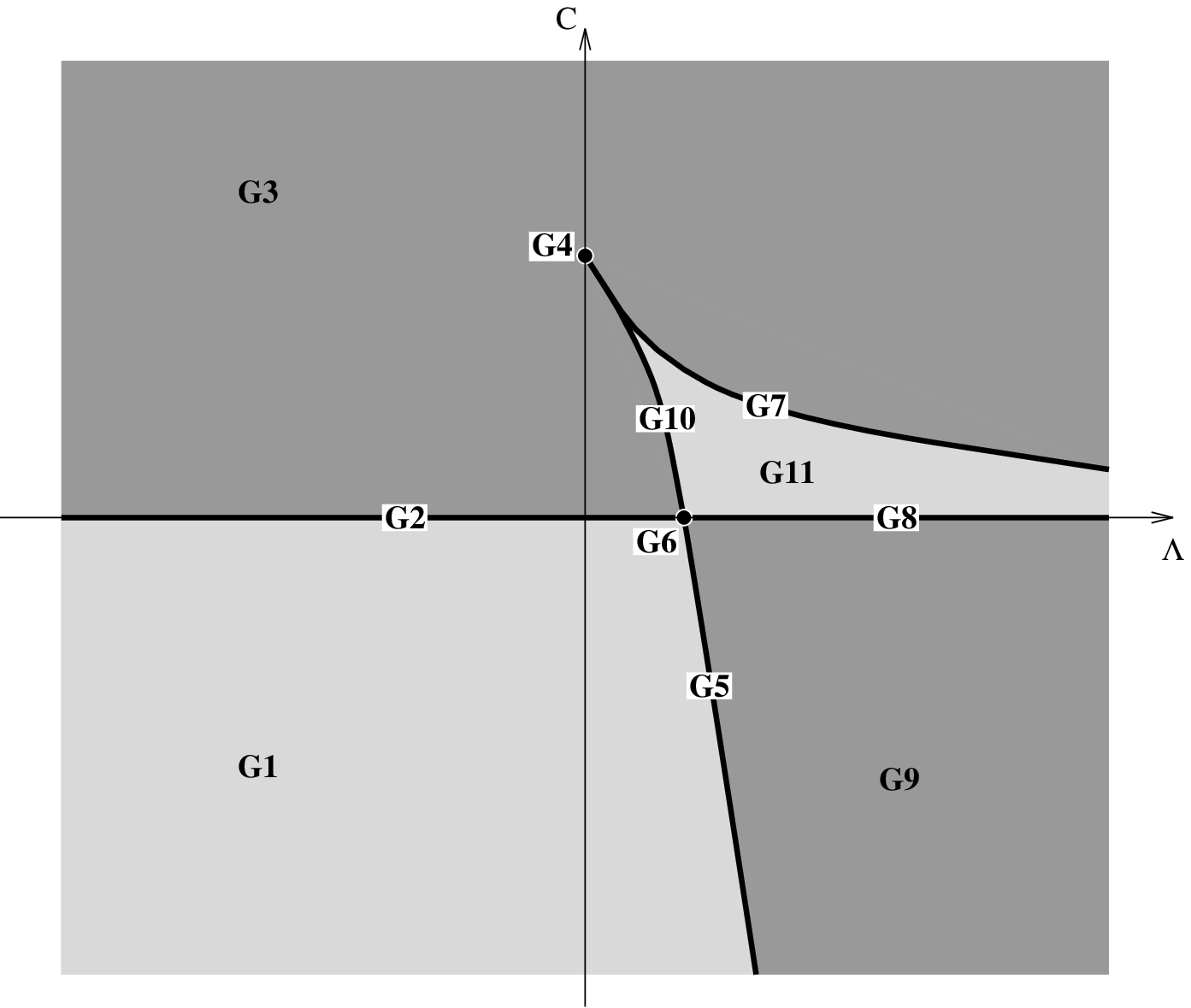}
\end{center}
\renewcommand{\baselinestretch}{.9}
\small \normalsize
{\bf Figure \Figtwelve:} {\small KV-model, overview.}
\end{figure}

This time we get 11 qualitatively different cases (in addition to two deSitter 
solutions) depending on the parameters
$C$ and $\Lambda$:

\smallskip
\begin{tabular}{l@{\ {\bf:}\quad}l}
{\bf G1,2}&no zeros of $h$ \\
{\bf G3}&one simple zero at $x^0=B$ \\
{\bf G4}&one triple zero at $x^0=1$ \\
{\bf G5,6}&one double zero at $x^0=e^{\sqrt\Lambda}$ \\
{\bf G7}&one double zero at $e^{-\sqrt\Lambda}$ and one simple zero at
                                                                    $B_1$ \\
{\bf G8,9}&two simple zeros at $B_2$ and $B_1$ \\
{\bf G10}&one simple zero at $B_3$ and one double zero at
                                                       $e^{\sqrt\Lambda}$ \\
{\bf G11}&three simple zeros at $B_3$, $B_2$, and $B_1$, \\
\end{tabular}

\smallskip
\noindent where $0<B_3<-\sqrt\Lambda<B_2<+\sqrt\Lambda<B_1$
and $B \in \dR^+$. An overview is provided by Fig.\ \Figtwelve. 
In the above list we took $x^0 \in \dR^+$.
The cases {\bf G2,6,8} and {\bf G1,5,9}, respectively, differ only in the
asymptotic behaviour of $h$ at $x^0\rightarrow 0$ ($h^{KV}\sim x^0$ for
$C\ne 0$, but $h^{KV}\sim (x^0\ln x^0)^2$ for $C=0$), which influences the
completeness properties (see below).
The critical values of $X^3$ are easily
determined to be $\pm \sqrt{\L}/\a$; the corresponding value of the
Casimir function $C$ is
\be C_{crit} = -4 \( \pm \sqrt\L -1 \)
  \exp\(\pm \sqrt{\L}\)  \, , \el CdeS
which marks the curve {\bf G5,6,10,4,7} of Fig.\ \Figtwelve\ and
simultaneously the deSitter solutions
$De^a=0, \, d\o = \pm {2\0\a}\sqrt{\L} \e$ (cf.\ Eq.\ \re{deSitter})).

\begin{figure}[t]%[h,t]
\begin{center}
\leavevmode
\epsfxsize 10cm \epsfbox{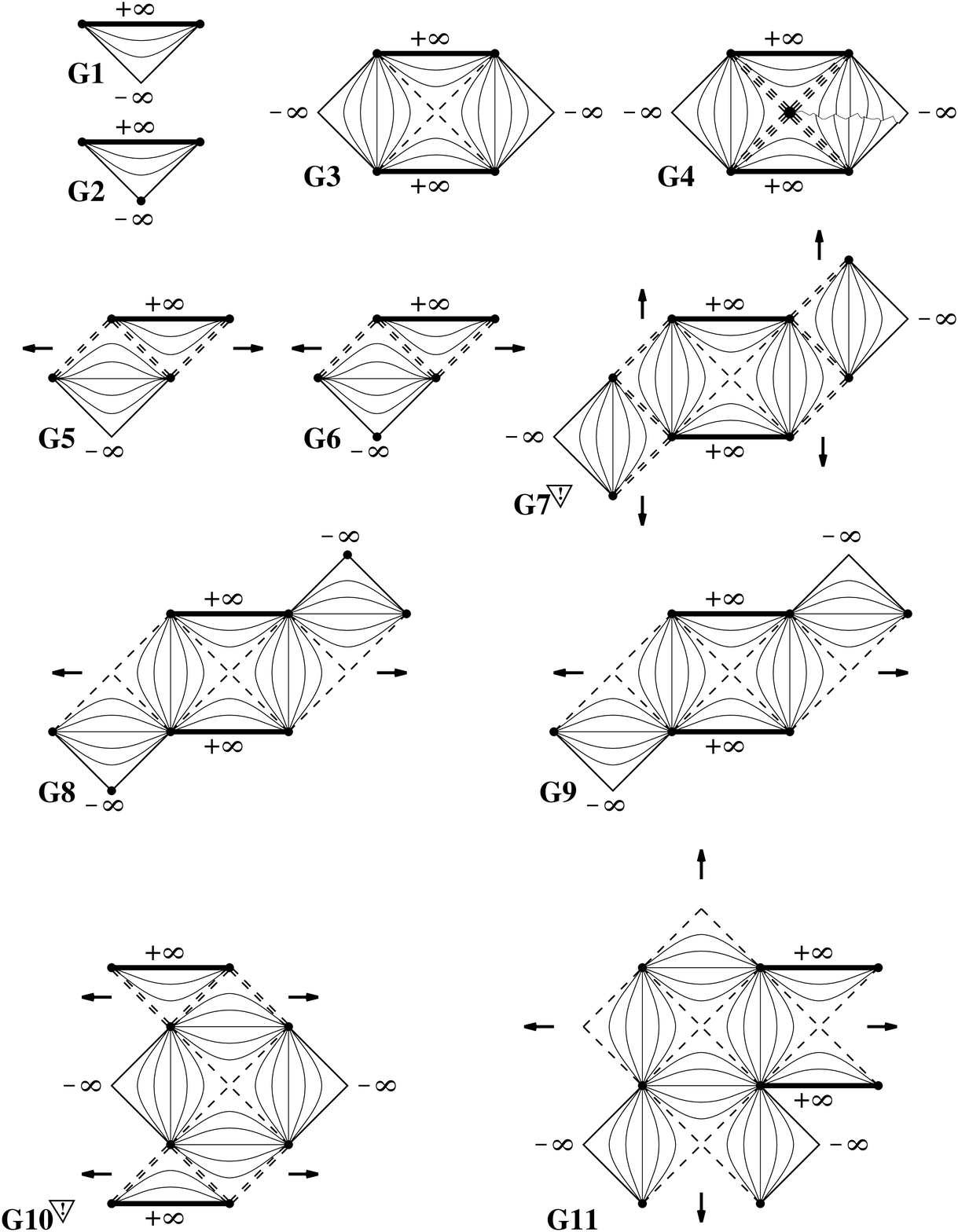}
\end{center}
\renewcommand{\baselinestretch}{.9}
\small \normalsize
{\bf Figure \Figthirteen:} {\small Penrose diagrams for the KV-model (those marked
\attn \ \,are non-smooth!).
Complete boundary lines
% (complete for all extremals)
are indicated by boldfaced lines, complete points (points at infinity)
by massive dots, incomplete boundary lines by thin solid lines,
and horizons by dashed lines (degenerate horizons, i.e.\ at higher order
zeros of $h$, by multiply dashed lines).
The null extremals \re{null1}, \ref{null2a})
which run through the diagrams under $\pm 45$ degrees are omitted.
Arrows outside a diagram ({\bf G5-11}) indicate that the solution
should be extended by appending similar copies at the corresponding
boundaries (but cf.\ Fig.\ 6e).}
\end{figure}

It is now straightforward to draw the Penrose diagrams of the KV-model.
The result for $\a>0$ is depicted in Fig.\ \Figthirteen\ (extended versions of
{\bf G5,9} are given in Fig.\ 14); the diagrams for $\a <0$ are
obtained by rotating these by $90$ degrees. According to the different
behaviour of $h$ at the boundary $x^0=0$ the Penrose diagrams
{\bf G2,6,8} ($C=0$) differ from  {\bf G1,5,9}, respectively, 
only by having a complete time-(space-)infinity, $\a>\!(<)\,0$, 
at that boundary (use \re{11})). Let us stress that all Penrose diagrams 
in Fig.\ \Figthirteen\ except for {\bf G7,10} are smooth now, the boundary lines
being actually straight (cf.\ footnote at the beginning of this section).
Again, however, a
representation of the extended diagrams for {\bf G4,7,10,11} will require
infinitely many overlapping layers (cf.\ Fig.\ 6e for the case {\bf G7}).

\begin{figure}[t]%[h,t]
\begin{center}
\leavevmode
\epsfxsize 12cm \epsfbox{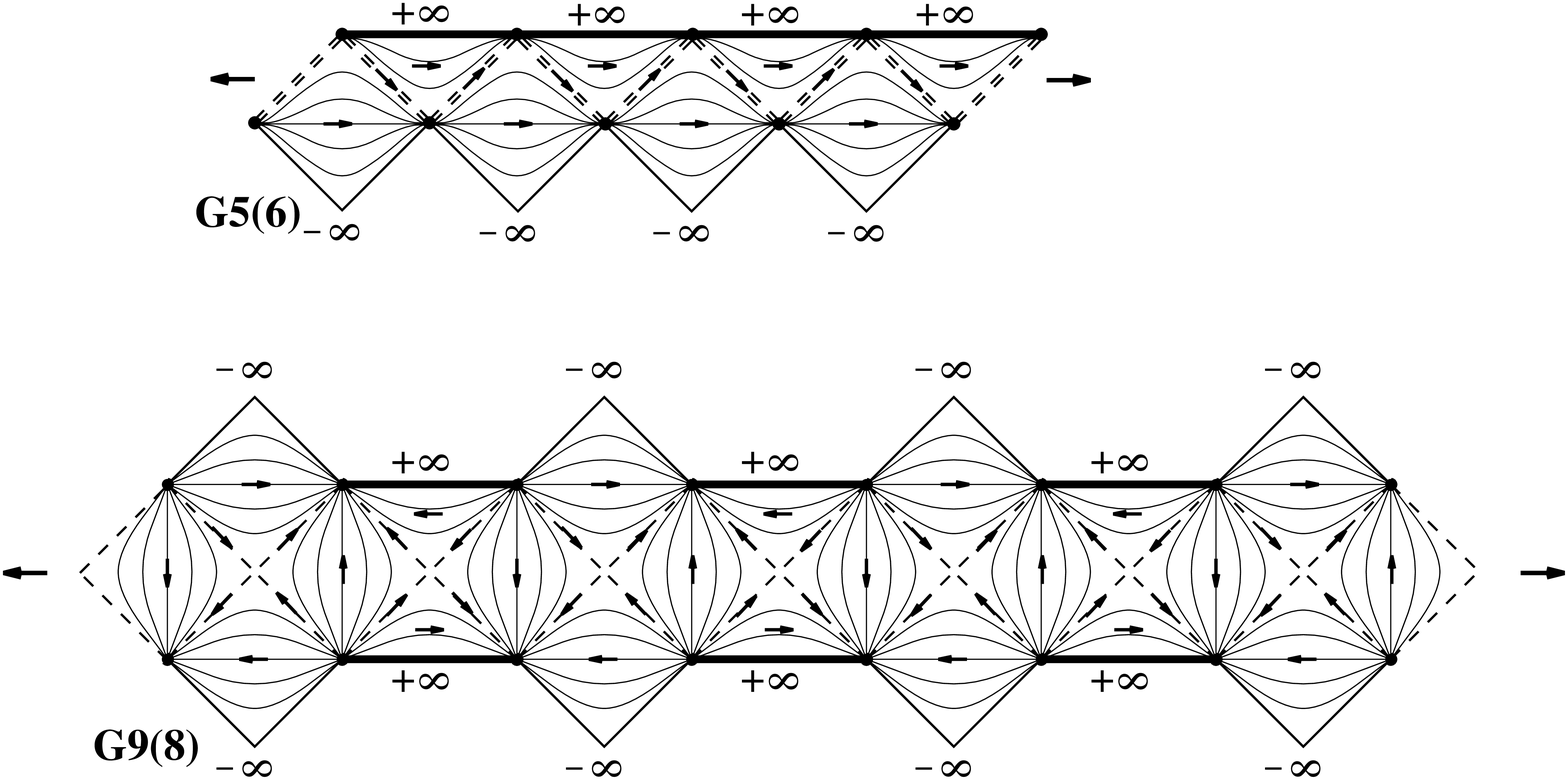}
\end{center}
\renewcommand{\baselinestretch}{.9}
\small \normalsize
{\bf Figure 14:} {\small Some extended Penrose diagrams for the KV-model.
Arrows inside the diagrams indicate Killing
fields.}
\end{figure}  

The numbering {\bf G1-11} has been chosen as in \pcite{Kat}, where the
Penrose diagrams Fig.\ \Figthirteen\ have been constructed first. It should be
noted, however, that our procedure to obtain these diagrams is incomparably
faster than the one of \pcite{Kat}. The main reason is that the local solutions
used there (resulting also from ours through the diffeomorphism \re{conf}))
are valid only in coordinate patches which are part of ours (the sectors
$h(x^0)\neq 0$); they had to be glued along their border, which entailed
lengthy considerations of the asymptotic behaviour. In the chiral gauge
\re{011h}), instead, the charts overlap and simply have to be matched
together.  As a consequence we also could prove that all the solutions of
\re{KV}), (and in fact also of \re{grav}) with an, e.g., analytic potential
$W$) are analytic. As pointed out repeatedly, the Penrose diagrams
{\bf G7,10} are, however, not smooth
but have to be regarded as schematic diagrams only (of course, the
universal covering solutions themselves {\em are\/} indeed smooth).

\medskip

Let us remark that for many of the Penrose diagrams, such as e.g.\ for
{\bf R1,G3,5,9}, it is possible to find also global coordinates displaying
explicitly the analyticity of $g$.  E.g., for {\bf R1,G3} we have already
found the Kruskal-like coordinates \re{Kruskal}) or the more explicit ones
\re{saddle}), and instead of for {\bf G5,9} we provided a global, explicitly
analytic chart in a chiral gauge \re{lcgauge}) for the (mathematically)
quite similar (four-dimensional) Reissner-Nordstr\"om solution in
\cite{rnletter}. The method presented there may be adapted easily to cases
such as {\bf G5,9} or in fact to any two-dimensional spacetime with a Killing
field that allows for a foliation by null-lines.

As a further example for the application of the simple rules outlined at
the beginning of this section we could consider, e.g., a model
\re{grav}) with $Z\equiv 0$ and $W=V(X^3)/2$ where 
$V$ is chosen to be the derivative of our generic function $h$ in Fig.\
2a. In this torsion-free case with, furthermore, $g=2e^+e^-$ [i.e.\
without a dilaton-dependent conformal factor, cf.\ Eq.\ \re{gconf})], the
function $h$ in \re{011h}) will  coincide with the function in Fig.\ 2a
up to a ``vertical'' shift  according to \re{Potential}).  It is now not
difficult to see that there will be 11 qualitatively different schematic
Penrose diagrams in this case, as well as further four deSitter solutions
(corresponding to the four distinct extrema of the function $h$ in Fig.\
2a or, equivalently, to the four zeros of the potential $V$). We leave it
as an exercise for the interested reader to sketch these Penrose diagrams
explicitly. Note that some of them will be  quite ramified already (cf.,
e.g., Fig.\ 5), a feature that becomes more and more pronounced with an
increasing number of zeros of $V$ (or $h$).

\section{Final Remarks and Outlook}

\plabel{Outlook}

Starting from the local solutions for the gravity models \re{grav})
and \re{gdil}), we have found all 
their maximally extended
universal covering solutions. They were found to be of an increasing
complexity with an increasing number of zeros of the potentials of
the Lagrangian.  Already in the simpler cases one obtains a diversity
of Penrose diagrams as was demonstrated, e.g., in Figs.\ 11 and \Figthirteen.
The corresponding solution space was found to split into several ``phases''
as illustrated in Figs.\ 10 and \Figtwelve\ 
(where, if $Z := const$, $\L$ may be identified
also with the ``Yang-Mills charge'' $q$, cf.\ the   
discussion around Eq.\ \re{poteff})).
Let us remark that within
figures of the latter type 
``phase transitions'' may be induced by 
scalar or fermionic matter fields: Given such a matter
distribution with a finite support  
in one null-coordinate on the space-time manifold, the
$C$- and $q$-values of the matter-free far past will in general
be changed into other ones for the matter-free far future (cf.\ also 
\cite{Heiko,CGHS}).

The next step, to be taken in Part III, will be to find all discrete
subgroups of the isometry groups of the universal coverings.  This
task is facilitated substantially by the fact that most solutions have
only one continuous symmetry (one Killing field), except for the
deSitter solutions, which have a maximal symmetry group (three
independent Killing fields). We will be able to classify and
parametrize all global, diffeomorphism inequivalent solutions of
arbitrary space-time topology.  Here the difference between the
various ``phases'' 
becomes quite pronounced: For instance, the only maximally extended
space-time solution of non-trivial topology with universal covering
{\bf G1} (cf.\ Figs.\ \Figtwelve,\Figthirteen) is cylindrical.
It will turn out to be a
two-parameter solution ($Z \equiv 0$); the second parameter besides $C$
will be found to be an appropriate metric-induced measure for the 
circumference of the cylinder.  A Penrose diagram of the type {\bf G11},
on the other hand, allows, through an identification of
opposite horizons in Fig.\ \Figthirteen\ (cf.\ arrows), for a spacetime topology
of a torus with hole.  Extending the diagram further into the vertical
direction by some blocks before identifying again all opposite
horizons, yields tori with arbitrarily many holes. These
(maximally extended) punctured torus solutions are parametrized by $2+h$ real
constants, the geometric origin of which shall be explained in Part
III; here $h$ denotes the number of holes ($h \ge 1$).  Other
identifications lead to punctured Riemann surfaces of higher genera,
too.  All of them are maximally extended, they have a non-trivial kink number
of the light-cone (just follow the movement of the light-cone along a loop
enclosing a hole), and the corresponding solution space has a dimension that
exceeds the number of generators of the fundamental group of the
respective space-time by one.

So, the global structure corresponding to a generic generalized dilaton theory
is much richer than that corresponding to ordinary (linear) dilaton gravity or,
e.g., spherically reduced vacuum gravity. 
This richness shows up both in the space-time solutions and in the
solution-space (= space of diffeomorphism inequivalent solutions).
Therefore, truly generalized dilaton theories allow one to address
qualitatively new aspects, e.g.:

\begin{itemize}
\item[---] May nontrivial space-time
topologies be described on the quantum level, given a quantum theory
constructed in a Hamiltonian framework (cf.\ Part IV) and limited as
such to the severely restricted topology-class $\S \times \dR$?
\item[---] How to quantize a solution-space of a highly non-trivial structure?
Or how to resolve the associated difficulties in a Dirac quantization
approach (cf.\ the appearance of ``winding numbers'' in the quantum states in
Part IV)?
\item[---] How are the quantum theories for Minkowskian and Euclidean signature
of the gravity theories related? 
\end{itemize} 
The last of these questions, for instance, is of current interest 
in the Ashtekar program of 4d quantum gravity \cite{Florenz}. We will find in
Part IV that while the spectrum of the ADM mass-operator coincides for both
signatures in the case of ordinary dilaton or spherically reduced gravity,
there are pronounced differences for other models in our general framework:
a continuous ADM mass-spectrum in the Minkowskian version, a discrete or
partially discrete one in the Euclidean counterpart. 

In an attempt to answer questions of the above kind difficulties arise
that are absent in simpler toy-models of gravity. They should  be faced,
however, in view of the incomparably more challenging task of quantizing
full 4d Einstein gravity. 

\section*{Acknowledgement}

We are grateful to H.\ Balasin, S.\ Lau, and H.\ Pelzer for discussions and
to H.D.\ Conradi, H.\ Kastrup, and W.\ Kummer for their interest in this work.
The work has been supported in part by the Austrian Fonds zur F\"orderung
der wissenschaftlichen Forschung (FWF), project P10221-PHY.

\end{document}